\documentclass[11pt]{article}
\pdfoutput=1
\usepackage{jheppub,amsmath,amssymb,mathabx}
\usepackage{enumerate}
\usepackage{bm}
\usepackage{bbm}
\usepackage[usenames,dvipsnames]{xcolor}
\usepackage{dsfont}
\usepackage{graphics}
\usepackage{tikz,todonotes}
\usepackage[utf8]{inputenc}
\usetikzlibrary{arrows,positioning,decorations.markings,decorations.pathmorphing,calc}

\definecolor{hyperref}{RGB}{026,028,087}
\allowdisplaybreaks[3]
\newcommand{\ep}{\epsilon}

\def\gsim{ \lower .75ex \hbox{$\sim$} \llap{\raise .27ex \hbox{$>$}} }
\def\lsim{ \lower .75ex \hbox{$\sim$} \llap{\raise .27ex \hbox{$<$}} }
\def\be{\begin{equation}}
\def\ee{\end{equation}}
\def\bea{\begin{eqnarray}}
\def\eea{\end{eqnarray}}
\newcommand{\nn}{\nonumber}

\newcommand{\Tc}{\mathcal{T}}

\def \bal#1\eal  {\begin{align} #1 \end{align}}
\newcommand{\ud} {\mathrm{d}}

\newcommand{\pd} {\partial}

\newcommand{\mcl} {\mathcal}
\newcommand{\tld}{\tilde}
\newcommand{\ai}{{\alpha}}
\newcommand{\bi}{{\beta}}

\newcommand{\li}{{\lambda}}
\newcommand{\ti}{{\tau}}

\newcommand{\ba}{\begin{array}}
\newcommand{\ea}{\end{array}}

\newcommand{\commentout}[1]{}

\newcommand{\comment}[1]{}

\newcommand{\bs}{\begin{split}}

\def\ba{\begin{eqnarray}}
\def\ea{\end{eqnarray}}
\newcommand{\F}[1]{{\mathcal F}\Bigg[{#1}\Bigg]}
\def\p{\partial}
\def\stu{St\"uckelberg }
\def\mpl{M_{\rm Pl}}

\def\nn{\nonumber}

\def\d{\mathrm{d}}

\def\mn{_{\mu \nu}}

\def\T{\mathcal{T}}

\def\stu{St\"uckelberg }
\def\Ost{Ostrogradsky }

\def\({\left(}
\def\){\right)}

\newcommand*{\mathcolor}{}
\def\mathcolor#1#{\mathcoloraux{#1}}
\newcommand*{\mathcoloraux}[3]{%
  \protect\leavevmode
  \begingroup
    \color#1{#2}#3%
  \endgroup
}
\newlength{\stheight}
\newcommand\textst[1][fu-grey]{
	\ifmmode\setlength{\stheight}{+1.0ex}
	\else\setlength{\stheight}{+0.5ex}
	\fi
	\bgroup\markoverwith{\textcolor{#1}{\rule[\the\stheight]{2pt}{1.0pt}}}\ULon
}
\newcommand{\textins}[2][fu-grey]{
	\ifmmode\mathcolor{#1}{#2}
	\else\textcolor{#1}{#2}\@\,
	\fi
}

\def\({\left(}
\def\){\right)}

\def\L{{\cal L}}

\begin{document}

\title{Positivity Bounds for Massive Spin-1 and Spin-2 Fields}

\author[a,b]{Claudia de Rham}
\author[a]{Scott Melville}
\author[a,b]{Andrew J. Tolley}
\author[a,c]{Shuang-Yong Zhou}
\affiliation[a]{Theoretical Physics, Blackett Laboratory, Imperial College, London, SW7 2AZ, U.K.}
\affiliation[b]{CERCAL, Department of Physics, Case Western Reserve University, 10900 Euclid Ave, Cleveland, OH 44106, USA}
\affiliation[c]{Interdisciplinary Center for Theoretical Study, University of Science and Technology of China, Hefted, Ansi 230026, China}

\emailAdd{c.de-rham@imperial.ac.uk}
\emailAdd{s.melville16@imperial.ac.uk}
\emailAdd{a.tolley@imperial.ac.uk}
\emailAdd{zhoushy@ustc.edu.cn}

\abstract{We apply the recently developed positivity bounds for particles with spin, applied away from the forward limit, to the low energy effective theories of massive spin-1 and spin-2 theories. For spin-1 theories, we consider the generic Proca EFT which arises at low energies from a heavy Higgs mechanism, and the special case of a charged Galileon for which the EFT is reorganized by the Galileon symmetry. For spin-2, we consider generic $\Lambda_5$ massive gravity theories and the special `ghost-free' $\Lambda_3$ theories.
Remarkably we find that at the level of 2-2 scattering,  the positivity bounds applied to $\Lambda_5$ massive gravity theories, impose the special tunings which generate the $\Lambda_3$ structure. For $\Lambda_3$ massive gravity theories, the island of positivity derived in the forward limit appears relatively stable against further bounds.  }

%\keywords{EFT, Unitarity, Positivity bounds, Transversity formalism}

\maketitle

%\newpage
%\setcounter{tocdepth}{2}
%\tableofcontents

%%%%%%%%%%%%%%%%%%%%%%%%%%%%
\section{Introduction}
\label{sec:intro}
%%%%%%%%%%%%%%%%%%%%%%%%%%%%

The powerful constraints of S-matrix analyticity, unitarity and crossing symmetry have been used to great effect in understanding the structure of Lorentz invariant theories at high energies, leading historically to the development of string theory. Some of these old ideas have been reinvigorated, in particular as part of the S-matrix bootstrap (and related conformal bootstrap) program (see for example \cite{Rattazzi:2008pe,Caron-Huot:2016icg,Paulos:2017fhb}).\\

In a related but distinct line of development, these constraints have also been brought to bear on the consistency of low energy effective theories, and imply `positivity bounds' on the signs of various combinations of coefficients of operators in the Wilsonian effective action.
The first such positivity bounds were developed previously as statements about forward limit scattering amplitudes for scalar particles in \cite{Adams:2006sv}. These forward limit bounds can also be applied to particles of arbitrary spin, including fermions, \cite{Bellazzini:2016xrt}, and previous applications of forward limit bounds to massive particles with spins include \cite{Cheung:2016yqr, Bonifacio:2016wcb}. Extending these bounds away from the forward limit spin zero particles was recently achieved in \cite{deRham:2017avq} (for earlier work see \cite{Pennington:1994kc,Vecchi:2007na,Manohar:2008tc,Nicolis:2009qm,Bellazzini:2014waa}) and have for example been applied to Galileon theories \cite{deRham:2017imi}, yielding new constraints on their parameter space. \\

The extension of this to all spins, including fermions, was achieved in \cite{deRham:2017zjm}. This development required developing dispersion relations for general spin scattering that have the same analyticity structure as scalar scattering, and the same positivity properties along both the left and right hand branch cut discontinuities away from the forward scattering limit. Previously it had only been known how to do this in special cases \cite{Mahoux:1969um}, and the key development of \cite{deRham:2017zjm} was to make use of the transversity formalism \cite{kotanski_transversity_1970,kotanski_kinematical_2016}, for which crossing symmetry for general spin particles becomes particularly simple. \\

In this article, we follow up the work of \cite{deRham:2017zjm} with two particularly important examples,  EFTs for massive spin-1 and spin-2 particles. The former has obvious relevance as the effective theory of massive gauge bosons in which the heavy Higgs fields which generate spontaneous symmetry breaking are integrated out. The latter are of relevance to massive gravity or multi-gravity theories and potentially the EFT for massive spin-2 mesons. The bounds will be applied under the assumption that these theories are weakly coupled in the sense defined in \cite{deRham:2017xox,Giudice:2007fh}. Specifically we will assume the existence of a weak coupling parameter $g_*$ which controls the scale of loop corrections $\hbar \sim g_*^2$ and is sufficiently small that we may apply the positivity bounds at tree-level. The precise validity of this requirement for the case of massive spin-2 (and spin-0) is discussed in more detail in \cite{deRham:2017xox}. Positivity requirements constrain the parameter space of these theories, and in particular the first $t$ derivative bounds are found to give an orthogonal constraint to the forward limit bounds alone. \\

The precise organization of the EFTs for each spin, depends somewhat on how they are anticipated to arise from some UV completion. In the case of spin-1, we consider two cases which we refer to as {\it Proca} and {\it Charged Galileon} defined as follows:

%%%%
\paragraph{Proca:}
In a Proca theory there are no underlying symmetries, other than the Poincar\'e and the $U(1)$ gauge symmetry which is spontaneously broken. The helicity-0 mode of the vector will then become dynamical and to elucidate this it is useful to introduce the \stu scalar field which describes this helicity-0 Goldstone mode, in such a way that the theory recovers a non-linearly realized gauge invariance. Such an EFT will arise through a Higgs mechanism, in which the Higgs particle is itself heavy, and may be integrated out, giving rise to an EFT for the \stu field. The interactions of the helicity-0 modes are typically stronger than those of the helicity-1 modes, and so the EFT is naturally organized as a Wilsonian EFT with an infinite number of local gauge invariant operators suppressed by a cutoff energy scale set by the helicity-0 interactions. \\

The construction of a generic EFT for such a field differs from the ghost-free Proca theory discussed in  \cite{Heisenberg:2014rta} because we allow higher derivative operators with the understanding that they are to be treated perturbatively, and that the would-be ghost they engender has a mass at or above the cutoff, meaning it does not exist. \\

%%%%
\paragraph{Charged Galileon:}
One particular tuning of the Proca theory that is of interest corresponds to the charged Galileon. This tuning guarantees that a Galileon symmetry emerges for the helicity-0 mode in the decoupling limit, and reproduces at leading order the ghost-free Proca theory of \cite{Heisenberg:2014rta}.
The physical interpretation of the fields is slightly different in this case. Now the \stu field (which transforms non-linearly under the gauge symmetry), can be thought of as the field of interest: a Galileon which is charged under the gauge symmetry and operators which explicitly break the Galileon symmetry will be naturally suppressed. \\

%%%%
\paragraph{Special Generalized Proca:}
The special class of Generalized Proca theories introduced in \cite{Heisenberg:2014rta} is built so as to ensure the absence of higher derivatives in equations of motion. This is of special interest if one is to consider for instance a Vainshtein-type of mechanism for the helicity-0 mode where we beyond the standard regime of validity of the LEEFT and consider a re-organization of the EFT as entering the Vainshtein region (see \cite{deRham:2014wfa} for a discussion of this point). This feature is particularly relevant in models of dark energy where the additional scalar degree of freedom is to be screened within the solar system. With this picture in mind, a special vector Galileon model was studied in the literature which only contains a cubic and quartic operator. \\[0.5cm]

In the case of spin-2 particles, we can imagine that the symmetry that is spontaneously broken by the mass term is either a linear spin-2 gauge symmetry, or is a nonlinear diffeomorphism symmetry. In the latter case, such a theory will be necessarily gravitational, and we shall focus on these massive gravity examples. Even within massive gravity, there is a choice in the organization of the EFT depending on what scale the cutoff of the EFT is assumed to be. A generic massive spin-2 field has a cutoff at the low scale $\Lambda_5=(m^4 \mpl)^{1/5}$. However by tuning the coefficients in the EFT, the scale may be raised to $\Lambda_3$ (but no higher in a local and Poincar\'e invariant theory, see for instance \cite{deRham:2016plk,Gabadadze:2017jom} for examples where the effective cutoff is raised to $\Lambda_2=(\mpl m)^{1/2}$).

%%%%
\paragraph{$\Lambda_5$ Massive Gravity:}
While the mass term for metric fluctuations softly breaks diffeomorphism invariance, which allows the helicity-1 and helicity-0 modes of the field to become dynamical, the \stu trick can again be used to introduce a scalar and vector field which non-linearly restore diffeomorphisms. The EFT is then naturally constructed in terms of these fields with the higher dimension operators in the Wilsonian effective action naturally suppressed by the scale\footnote{Strictly speaking $\Lambda_5 = (m^4 M)^{1/5}$ where $M^2=\mpl^2 g_*^2$ and $g_*$ is the weak coupling parameter to be introduced later \cite{deRham:2017xox}.} $\Lambda_5 = (m^4 \mpl)^{1/5}$. In this case we find that the positivity requirements seem to necessitate a tuning of the coefficients which raises the cutoff from $\Lambda_5$ to $\Lambda_3$---at least for those interactions which contribution to the four-point amplitude at tree-level. \\

%%%%
\paragraph{$\Lambda_3$ Massive Gravity:}
As is now well known, there is a further tuning which one can perform which can be phenomenologically relevant because it raises the cutoff to $\Lambda_3 = (m^2 \mpl )^{1/3}$ \cite{deRham:2010ik,deRham:2010kj}\footnote{More generally $\Lambda_3 = (m^2 M )^{1/3}$  with  $M^2=\mpl^2 g_*^2$. }. This is the special ghost-free massive gravity theory, although from an weakly coupled UV completion EFT point of view \cite{deRham:2017xox} these are only the leading terms in an infinite expansion of irrelevant operators. This theory is known to satisfy the forward limit constraints in a finite region of parameter space \cite{Cheung:2016yqr}, and here we show that this region is (mostly) consistent with the higher $t$ derivative bounds also---a promising sign that local weakly coupled UV completion of massive gravity may be possible. \\[0.5cm]

To begin, we briefly review the positivity bounds in Section \ref{sec:positivity}, both at and away from the forward limit, and describe a convenient basis of crossing symmetric variables. Then in Sections~\ref{sec:proca} and~\ref{sec:galileon}, we construct the most general EFTs of a massive spin-1 (Proca) field and of a charged Galileon from the bottom-up, and show how the parameter space is constrained by the positivity bounds. We then focus on a special simple example of Vector Galileon in Section~\ref{sec:SimplestVec} and show how the positivity constraints require the presence of additional operators.
Next, we present in Section~\ref{sec:L5} the analogous bottom-up construction for a massive spin-2 field (massive gravity), and show how forward and $t$ derivative bounds effectively impose the special tunings which generate $\Lambda_3$ massive gravity, as far as tree-level 2-2 scattering amplitudes are concerned. Finally, in Section~\ref{sec:L3}, we focus on the $\Lambda_3$ theory, and confirm that $\Lambda_3$ massive gravity possesses a non-trivial region of parameter in which local, Lorentz invariant UV completion may be possible, even when faced with new $t$ derivative constraints. In the conclusion we briefly discuss the UV completion and possible existence of a Higgs mechanism for spin 2 particles.

%%%%%%%%%%%%%%%%%%%%%%%%%%%%
\section{Review of Positivity Bounds}
\label{sec:positivity}
%%%%%%%%%%%%%%%%%%%%%%%%%%%%

In this section we briefly review the relevant details of the general positivity bounds for scalar field \cite{deRham:2017avq} and for arbitrary spin scattering which is discussed at length in Ref.~\cite{deRham:2017zjm}. Those positivity bounds rely on assumptions that the high energy completion of the  low-energy effective field theory we are interested in is Lorentz-invariant, local and analytic. The positivity bounds we shall rederive below are bounds on the properties of the whole non-perturbative 2-2 elastic scattering amplitude and make no assumption on whether or not the UV completion is weakly coupled. However in practise the most straightforward way to apply those bounds is to assume a weakly coupled UV completion and compute the scattering amplitude perturbatively. The precise nature of this weak coupling assumption as well as possible alternatives are summarized below.

\subsection{Weak Coupling and Alternatives}

The positivity bounds \eqref{eqn:forwardbounds} and \eqref{eqn:forwardbounds_definite} that we shall derive below must hold for \emph{any} local, Lorentz-invariant and analytic Wilsonian UV completion. However  in practice to apply them  concretely, one must have a way to reliably compute the relevant EFT scattering amplitudes and we shall discuss more precisely how to proceed below.

\paragraph{Weakly Coupled UV Completion:} In this work, we shall consider various low energy Lagrangians with the implicit assumption that loops are suppressed by a small coupling, $g_*$, so that amplitudes can be calculated by straightforward perturbation theory and we shall only apply the positivity constraints to the tree level part of the EFT amplitudes. Focusing on the tree-level amplitude relies on the assumption that some version of weak coupling persists in the UV, and hence prevents heavy loops from entering the dispersion relation and spoiling the above arguments---see \citep{deRham:2017xox} for a careful discussion of the interplay between different loop orders mandated by analyticity.
One interesting aspect that occurs when applying the positivity bounds to tree-level amplitudes assuming a weakly coupled UV completion is that the imaginary part of the 2-2 scattering amplitude does not start when the center of mass energy of the amplitude is $2m$ but rather when it reaches the scale of the next massive particle beyond the regime of validity of the EFT, hence making the positivity bounds more constraining.

\paragraph{Alternatives to Weakly Coupled UV Completion:} We emphasize again that the weak coupling assumption is a standard but yet purely technical one in applying the positivity bounds. Another way to make progress is to take into account loop contributions from light fields. This is discussed explicitly in \cite{deRham:2017xox}. In this approach we compute the scattering amplitude in the low energy effective theory to any desired order in the loop expansion. We can then calculate the imaginary part that arises from the light loops at low energies. We may then subtract from the dispersion relation this calculable contribution with the dispersion integral cutoff at some scale $\epsilon \Lambda$ where $\Lambda$ is the strong coupling scale and $\epsilon$ is taken sufficiently small that we can trust the loop expansion. This results in a new dispersion relation in which the branch cuts begin at $s = ( \epsilon \Lambda)^2$, which will nevertheless lead to positivity. In this approach there is no requirement to assume that the UV completion is weakly coupled. The only assumption is that the low energy effective theory does a good job at capturing the low energy physics, as it should.

\paragraph{Other Approaches:} In this manuscript we shall follow the logic that the low-energy EFT should satisfy the positivity bounds by itself and apply constraints on the spin-1 and spin-2 EFTs with this consideration in mind. It is however worth pointing out that alternatives do exist. First, the coupling of the EFT to new light fields may play a crucial role in preserving the positivity bounds. Alternatively, the assumptions underlying the positivity bounds may themselves be questioned. In this sense, any region of parameter space that we shall rule out by the positivity bounds in what follows would require the existence of a `nonstandard' high energy completion to be viable. This `nonstandard' UV completion may possibly be as `benign' as breaking some notion of strict locality at arbitrarily high energy (possibly even above the Planck scale) or could be much more severe (eg. breaking causality). By themselves the failure of satisfying positivity bounds does not determine how `dramatic' the required UV completion would be, it only states that it is not standard (in the sense imposed by analyticity and strict locality).

\subsection{From Scalar to Spinning Positivity Bounds}

In what follows we shall consider a 2-2 scattering amplitude with Mandelstam variables $s, t, u$. 
The first positivity bounds for scalar fields were provided and used in the forward limit ($t=0$) in \cite{Adams:2006sv} although their validity beyond the forward limit was suggested in \cite{Pennington:1994kc,Vecchi:2007na,Manohar:2008tc,Nicolis:2009qm,Bellazzini:2014waa}.\\

For scalar fields, it is indeed  relatively straightforward to derive all the scalar positivity bounds away from the forward limit once the positivity of the $t$ derivatives of the imaginary part of the amplitude away from the forward limit is established and we refer to Ref.~\cite{deRham:2017avq} for the precise form of those bounds. However, subtleties arise when generalizing these bounds to nonzero spins  \cite{deRham:2017zjm}. This is because spin structures introduce kinematic singularities in the scattering amplitude on top of the physical poles and branch cuts and the crossing relations are non-trivial in the usual helicity formalism. Fortunately, crossing relations can be diagonalized in the so-called transversity formalism \cite{Kotanski:1965zz} where the spin quantization axis is chosen to be orthogonal to the interaction plane, and the kinematic singularities can be taken care by studying a modified amplitude, with extra subtractions \cite{deRham:2017zjm}. For a recent application of the transversity formalism see \cite{Bonifacio:2018vzv}. \\

In the following, we shall summarize the main results for the special case of scattering of {\it four bosons with the same mass $m$ and spin-$S$}, and an interested reader is referred to \cite{deRham:2017zjm} for more details and for the cases with different masses and/or spins including fermions. We note in particular that the formalism naturally accommodates for scatterings of particles with different masses and spins, but to simplify the presentation here we shall on same mass and spin.

\paragraph{Crossing Relation:} Unitarity implies that the imaginary part of the scattering amplitude is positive in the physical $s>4m^2$ region. Through a Cauchy integral, analyticity then relates the imaginary part of this amplitude to the Mandelstam region and in doing so also requires knowledge of the imaginary part of the $u$-channel amplitude.
To this end, a crucial ingredient in deriving positivity bounds from a dispersion relation is the \emph{crossing relation} which identifies $s$- and $u$- channel amplitudes. In particular, the exchange of two particles corresponds to a trivial reordering of the annihilation/creation operators within an $S$-matrix element, so that amplitudes are related,
\begin{equation}
\mathcal{A}^{A+B \to C+D} ( p_1, p_2, p_3, p_4 ) = \eta \mathcal{A}^{A+\bar{D} \to C+\bar{B}} ( p_1, -p_4, p_3, -p_2 )
\end{equation}
where $\eta$ is an overall sign determined by the statistics of the particles, and $\bar{B}$ denotes the antiparticle of $B$. Note that the momenta have also been reordered. To return to the standard configuration $\{ p_1, p_2, p_3, p_4 \}$ which one uses in computing the amplitude, one must apply a Lorentz transformation.

For scalar particles, this Lorentz transformation acts trivially on the amplitude, and consequently the crossing relation is straightforward,
\begin{equation}
\mathcal{A}^{A+B \to C+D} ( s, t ) = \mathcal{A}^{A+\bar{D} \to C+\bar{B}} ( u, t )  \;\;\;\; \text{ for scalars.}
\end{equation}
For particles with spin however, the action of the Lorentz transformation is generally rather complicated (and not sign definite).

In the helicity basis of polarizations, which quantize spin along the particle's momenta, a single configuration of helicities is mapped under crossing to a sum over all available configurations, which is not sign definite. This prevents one from deriving positivity bounds for helicity amplitudes\footnote{Except for the special cases in which $m=0$, or the forward limit $t=0$, for which the crossing relation becomes trivial.}.
Instead, one can quantize the spins of all particles along the \emph{normal} to the scattering plane, known as the transversity basis. This has the advantage that under crossing symmetry, a single configuration of transversities is mapped to a single configuration of transversities, preserving positivity properties.

\paragraph{Transversity amplitudes:}
The transversity formalism is simply a change of basis from the usual helicity formalism for scattering amplitudes that serves to diagonalize the crossing relations. 
At high energies, we recover that the behaviour of the transversity amplitude is governed by the \stu fields as we shall see for instance in \eqref{eq:highenergy}.

In what follows, we shall  characterize external states in terms of transversity eigenstates where the transversity is the Pauli-Lubankski four vector projected onto the normal to the scattering plane. The transversity eigenstates can be obtained from the helicity eigenstate by a Wigner rotation $u^S_{\li \tau} = D^S_{\li \tau}(\pi/2, \pi/2,-\pi/2)$, where $D^S$ is the usual Wigner $D$-matrix associated with rotation $e^{-i\pi J_z/2}e^{-i\pi J_y/2}e^{i\pi J_z/2}$. Explicitly the transversity amplitude $\mcl{T}_{\ti_1 \ti_2 \ti_3 \ti_4}$ is given in terms of the helicity amplitude $\mcl{H}_{\li_1 \li_2 \li_3 \li_4}$ via
\be
\mcl{T}_{\ti_1 \ti_2 \ti_3 \ti_4} =  \sum_{\li_1, \li_2, \li_3, \li_4} u^S_{\li_1\ti_1} u^S_{\li_2\ti_2} u^{S*}_{\ti_3\li_3} u^{S*}_{\ti_4\li_4} \mcl{H}_{\li_1\li_2\li_3\li_4}   ,
\ee
where ${}^*$ is the complex conjugate. In the transversity basis, the crossing relation is diagonal, for example, in $s-u$ crossing, if $\mcl{T}^s_{\ti_1 \ti_2 \ti_3 \ti_4}(s,t,u)$ denotes the $s$-channel process $A+B \rightarrow C+D$ and $\mcl{T}^u_{\ti_1 \ti_2 \ti_3 \ti_4}(s,t,u)$ the $u$-channel process $A+\bar D \rightarrow B+\bar C$, then crossing symmetry states that \cite{deRham:2017zjm}:
\be
\mcl{T}^s_{\ti_1 \ti_2 \ti_3 \ti_4}(s,t,u)  = e^{i\pi\sum_i \ti_i} e^{-i \chi_u \sum_i \ti_i}  \mcl{T}^u_{-\ti_1 -\ti_4 -\ti_3 -\ti_2}(u,t,s)    ,
\ee
where $\exp({\pm i \chi_u}) = (-su\mp 2im\sqrt{stu})/\sqrt{s(s-4m^2)u(u-4m^2)}$. For any process $\mcl{T}_{\ti_1 \ti_2 \ti_3 \ti_4}$ contains unphysical kinematic singularities, so we instead consider the modified amplitude
\be
\mcl{T}^+_{\ti_1 \ti_2 \ti_3 \ti_4}(s,\theta)  =   (s(s-4m^2))^{N_S/2 - 1}  \left(  \mcl{T}_{\ti_1 \ti_2 \ti_3 \ti_4}(s,\theta) + \mcl{T}_{\ti_1 \ti_2 \ti_3 \ti_4}(s,-\theta) \right)\,,
\ee
where $\theta$ is the scattering angle in the center of mass frame, and $N_S$ is a sufficiently large integer, $N_S \geq 2 + 2| \tau_1 + \tau_2 |$
that will determine the number of subtractions needed in the dispersion relation. Then one can show that the $t$ derivatives of the imaginary part of an elastic amplitude $\mcl{T}^+_{\ti_1 \ti_2 \ti_1 \ti_2}$ are both positive on the left and right hand branch cuts of the $s$ complex plane away from the forward limit ($0\le t<m^2$), which puts $\mcl{T}^+_{\ti_1 \ti_2 \ti_1 \ti_2}$ on the same footing as a scalar amplitude.  When considering the forward limit we may take superpositions of transversity states for which the number of subtractions must be taken to be at least $N_S = 2+ 2(S_1+S_2)$ to be valid for all combinations of transversities.

\paragraph{Dispersion Relation:}
The pole subtracted $\tilde{\mcl{T}}^+_{\ti_1 \ti_2 \ti_1 \ti_2}$ obeys a dispersion relation analogous to the scalar case. Performing $N_S$ subtractions to ensure that an integration contour in the complex $s$ plane can be closed at infinity\footnote{
From the Froissart-Martin bound applied even away from the forward limit $0 \le t < m^2$, the asymptotic growth of ${\mcl{T}}^+_{\ti_1 \ti_2 \ti_1 \ti_2}$ is at most $s^{N_S}$.
},
\be
\label{fvtdef}
f_{\ti_1\ti_2}(v,t) =  \frac{1}{N_S!} \frac{\pd^{N_S}}{\pd s^{N_S}}  {\tilde{\mcl{T}}}^+_{\ti_1\ti_2\ti_1\ti_2}(s\!=\!2m^2-t/2+v,t)\,,
\ee
we have the dispersion relation,
\ba
\label{faltdef2}
 f_{\tau_1 \tau_2 } (v, t)
 =  \frac{1}{\pi}  \int_{\mu_b}^\infty \ud \mu  \frac{ {\text{Abs}}_s \Tc^+_{\tau_1 \tau_2 \tau_1 \tau_2}(\mu,t) }{  (\mu-2m^2+t/2 - v)^{N_S+1 } }+    \frac{1}{\pi}  \int_{\mu_b}^\infty \ud \mu  \frac{  {\text{Abs}}_u \Tc^+_{\tau_1 \tau_2 \tau_1 \tau_2}(4m^2-t-\mu,t) }{ ( \mu -2m^2+t/2+v)^{N_S+1} } \, ,\qquad
 \ea
where $\mu_b$ is the scale at which the branch cut begins---in general $\mu_b = 4m^2$ from light loops, but when considering tree-level scattering $\mu_b$ can be taken as $\Lambda_{\rm th}^2$, where $\Lambda_{\rm th}$ is the mass of the first new heavy degrees of freedom outside the EFT.

\paragraph{Positivity Bounds:}
Manipulating \eqref{faltdef2}, the following positivity bounds can be established,
\bal
\label{eq:2ndbound}
&f_{\ti_1\ti_2}(v,t) >0,
\\
&\frac{\pd}{\pd t} f_{\ti_1\ti_2}(v,t)  + \frac{N_S+1}{2\mcl{M}^2} f_{\ti_1\ti_2}(v,t)  >0 ,
\\
&\frac12 \frac{\pd^2}{\pd t^2} f_{\ti_1\ti_2}( v,t) + \frac{N_S+1}{2\mcl{M}^2} \left( \frac{\pd}{\pd t} f_{\ti_1\ti_2}( v,t)  + \frac{N_S+1}{2\mcl{M}^2} f_{\ti_1\ti_2}( v,t) \right)  >0  ,
\eal
for any $\{ v, t , \mathcal{M} \}$ in the ranges,
\be
|v | < \mu_b - 2m^2 + t/2 , \;\;\;\; 0\le t < m^2 ,\;\;\;\; \mathcal{M}^2 \leq \mu_b - 2m^2 + t/2 \pm v ,
\label{eqn:vtM}
\ee
i.e. any $v$ within the radius defined by the branch cuts, any positive $t$ up to the first singularity (which in this case is a pole at $m^2$, but in other cases could be as large as $\mu_b$), and any $\mathcal{M}^2$ which is sufficiently small to ensure cancellation of negative terms arising from $t$ derivatives acting on the denominator of the integrand.  \\

In fact, following the procedure described in \cite{deRham:2017zjm}, one can derive an infinite number of positivity bounds,
\be
Y_{\ti_1\ti_2}^{(2N,M)} (t) >0
\ee
with $Y_{\ti_1\ti_2}^{(2N,M)}$ defined by the recursion relation
\bal
Y_{\ti_1\ti_2}^{(2N,M)} (t) & =  \sum_{r=0}^{M/2} c_r B_{\ti_1\ti_2}^{(2N+2r,M-2r)}(t)
\nn\\
& + \frac{1}{\mcl{M}^2} \sum^{(M-1)/2}_{{\rm even}~k=0} (2N+2k+1)\beta_k Y_{\ti_1\ti_2}^{(2N+2k,M-2k-1)}(t)    ,
\eal
where
\be
B_{\ti_1\ti_2}^{(2N,M)}(t)\equiv \frac1{M!}  \frac{\pd^{2N}}{\pd v^{2N}} \frac{\pd^M}{\pd t^M} \tld{\mcl{T}}^+_{\ti_1\ti_2\ti_1\ti_2}(s\!=\!2m^2-t/2+v,t)   ,~~~  \forall~~ N\ge \frac{N_S}{2},~M>0
\ee
where $N_S \geq 2 + | \tau_1 + \tau_2 |$, $\{v,t,\mathcal{M}^2 \}$ lie in the ranges \eqref{eqn:vtM}, and $c_r$ and $\beta_k$ are defined by
\be
c_k = - \sum^{k-1}_{r=0} \frac{2^{2(r-k)}c_r}{(2k-2r)!} ~~{\rm and}~~ \beta_k = (-1)^k \sum^k_{r=0} \frac{2^{2(r-k)-1}c_r}{(2k-2r+1)!}, ~~k\ge 1    ,
\ee
($c_0\equiv 1$ and $Y_{\ti_1\ti_2}^{(2N,0)}\equiv B_{\ti_1\ti_2}^{(2N,0)})$.  \\

The scale $\mathcal{M}^2$ must be at least $\sim 2m^2$, but can be significantly larger if we restrict our attention to weakly coupled UV completions. In this case, tree-level amplitudes are good approximations to the full amplitudes up to some threshold scale $\Lambda_{\rm th}$, i.e. the cutoff of the EFT,  and one can take $\mathcal{M}^2 \sim \Lambda_{\rm th}^2$, which significantly improves the efficiency of the positivity bounds and is the assumption we will make in the following. Alternatively, one can subtract the contribution of light loops to the amplitude within the EFT regime up to some sub-threshold scale $\epsilon \Lambda_{\rm th}$, and then $(\epsilon \Lambda_{\rm th})^2$ sets the scale of $\mathcal{M}^2$ \cite{deRham:2017imi,deRham:2017xox}. \\

\paragraph{Indefinite Transversity Bounds:}
One can also construct a $f_{\ai\bi}(0,t)$ for mixed (indefinite) transversity states ($\ai,\bi$):
\be
f_{\ai\bi}(0,t) = \sum_{\ti_1, \ti_2,\ti_3, \ti_4}  \ai_{\ti_1}\bi_{\ti_2} \ai_{\ti_3}^* \bi_{\ti_4}^* f_{\ti_1\ti_2\ti_3\ti_4}(0,t)   ,
\ee
where $ f_{\ti_1\ti_2\ti_3\ti_4}$ is defined analogous to $f_{\ti_1\ti_2}$ but not restricting to elastic processes. Then it can be shown that, in the forward limit, we have for the indefinite states,
\be
\label{eqn:forwardbounds}
 \frac{\pd^{2N}}{\pd v^{2N} } f_{\ai\bi}(0,0) >0, ~~~\forall ~~~N  \,,
\ee
while for the definite states, the bound is also preserves away from the forward limit
\ba
 \frac{\pd^{2N}}{\pd v^{2N} } f^{\rm definite}_{\ai\bi}(0,t) >0, ~~~\forall ~~~N \quad{\rm and }\quad 0\le t <m^2  \,.
\label{eqn:forwardbounds_definite}
\ea
When true for indefinite transversity states, it is by extension true for both definite and indefinite helicity states. The same is not true for the $t$-derivative bounds which have for example no simple analogue in the helicity basis. In what follows it will be useful to use as a shorthand notation the quantity $f_{\ai\bi}(0,t)$ even when discussing different definite transversity bounds: since the definite transversity positivity bounds are the coefficients of the expansion in $\mcl{O}(\ai_{\ti_1}\bi_{\ti_2})^2$.

\paragraph{Number of Subtractions:}

To reiterate, while $N_S = 2 + 2 | \tau_1 + \tau_2 |$
is a necessary number of subtractions for a given definite transversity scattering process, for $S_1=S_2=S$, $N_S = 2 + 4 S$ is always sufficient \emph{for any polarization}. When scattering indefinite combinations of transversities, it is this latter value of $N_S$ which we shall use. For uniformity of presentation, we shall always quote values of $f_{\tau_1 \tau_2 }$ calculated using this sufficient $N_S$, regardless of $| \tau_1 + \tau_2 |$. In practice, performing only the necessary subtractions does not change the qualitative form of the bounds, only some numerical prefactors.  Given this some of our quoted results are not necessarily the tightest form of the bounds, however the differences are found to be small and do not affect the qualitative picture.

%
%\paragraph{Weakly Coupled UV Completion:}
%The positivity bounds \eqref{eqn:forwardbounds} and \eqref{eqn:forwardbounds_definite} must hold for \emph{any} local, Lorentz-invariant and analytic Wilsonian UV completion. However  in practice to apply them  concretely, one must have a way to reliably compute the relevant EFT scattering amplitudes. In this work, we shall consider various low energy Lagrangians with the implicit assumption that loops are suppressed by a small coupling, $g_*$, so that amplitudes can be calculated by straightforward perturbation theory. Moreover, in order to apply positivity constraints to only the tree level part of the EFT amplitude, one must assume that some version of this weak coupling persists in the UV, to prevent heavy loops from entering the dispersion relation and spoiling the above arguments---see \citep{deRham:2017xox} for a careful discussion of the interplay between different loop orders mandated by analyticity. 

%%%%%%%%%%%%%%%%%%%%%%%%%%%%
\section{Massive Spin-1 Field}
\label{sec:proca}
%%%%%%%%%%%%%%%%%%%%%%%%%%%%

Having reviewed the framework, we now apply it to massive spin-1 fields and  construct a generic massive spin-1 effective field theory from the bottom up, before deriving constraints on the parameter space from the positivity bounds assuming the existence of a standard UV completion.

\subsection{Symmetry Breaking}
\label{sec:scales}

 It is useful to first start the discussion with a single massless spin-1 field $A_\mu$. Then $U(1)$ gauge invariance implies that the EFT description of such a theory is built out the Maxwell tensor $F\mn$ and its derivatives. Assuming an EFT with a cutoff $\Lambda_A$ for a massless spin-1 field, we then ought to consider all operators of the form,
 \ba
 \label{eq:LA}
 \L_A= \frac{\Lambda_A^4}{g_*^2}\, \F{\frac{\p}{\Lambda_A},\frac{F}{\Lambda_A^2}}\,,
 \ea
where $\mathcal F$ designates a dimensionless Lorentz scalar formed from its arguments, and each operator is multiplied by an \emph{a priori} undetermined Wilson coefficient. The overall scale $\Lambda_A$ has been chosen so that the individual coefficients are each order unity (or smaller)---this ensures that perturbative unitarity is obeyed on all scales sufficiently below $\Lambda_A$ (for which $\partial/\Lambda_A \ll 1$ suppresses the non-renormalizable operators). If the overall coupling $g_* \sim 1$ then the theory would be expected to be strongly coupled near $\Lambda_A$. In this case it would be necessary to use the improved positivity bounds and subtract off the light loops \cite{deRham:2017imi,deRham:2017xox}. If $g_* \ll 1$ then these loop contributions will be small, since $g_*^2$ counts the number of loops, and we may apply the positivity bounds at tree level.  \\

Now considering a symmetry breaking scheme that may occur at a different scale and generates new operators that involve the helicity-0 mode $\phi$ of the now massive spin-1 field. Then in addition to the operators considered in  EFT of such a theory \eqref{eq:LA}, the EFT also involves operators built out of the  ``covariant\footnote{
Note that the notation $D_\mu \phi$ should not be confused with the usual fundamental/adjoint gauge covariant derivative; Here $\phi$ transforms \emph{non-linearly} under the gauge symmetry.
}" term
\ba
\phi_\mu=D_\mu \phi = \partial_\mu \phi + m A_\mu\,,
\ea
leading to an additional sector of the form
\ba
\label{eq:Lphi}
\L_\phi=\frac{\Lambda_\phi^4}{g_*^2} \, \F{\frac{\p}{\Lambda_\phi},\frac{\phi_\mu}{\Lambda_\phi^2}}\,,
\ea
where the cutoff scale $\Lambda_\phi$ for the symmetry breaking sector may be independent to $\Lambda_A$ (note that for the EFT description to make sense we require $m \ll \Lambda_{A,\phi}$). Quantum corrections generically cause the scales to flow between different operators, so care must be taken if one is to ensure that the Wilson coefficients are reliably of order unity. For instance, since $(\partial_{\mu} \phi_{\nu}-\partial_{\nu} \phi_{\mu}) = m F_{\mu\nu}$ the symmetry breaking sector generates interactions of the form (schematically)
\ba
g_*^2  \Delta \mathcal{L}_{\phi} \supset \Lambda_{\phi}^4 \left( \frac{D \phi}{ \Lambda_{\phi}^2}   \right)^{2} \left( \frac{m F}{ \Lambda_{\phi}^3}   \right)^{2}+ \Lambda_{\phi}^4 \left( \frac{m F}{ \Lambda_{\phi}^3}   \right)^{4}
\sim  \left( \frac{m^2}{\Lambda_{\phi}^6 } \right)  F^{2} (D \phi)^2 +\frac{m^4}{\Lambda_{\phi}^8}F^4 \,.
\ea
Although the cutoff in the symmetry breaking sector is $\Lambda_{\phi}$ we do not generate pure gauge interactions at $\Lambda_{\phi}$ at this order due to the fact that in the limit $m \rightarrow 0$ the two sectors decouple. This property is preserved under loops of the light field. This is easily shown since in dimensional regularization all diverging momentum integrals are replaced by positive powers of $m$ and thus higher loop corrections are $m/\Lambda_{\phi}$ suppressed\footnote{In our scheme, the Wilsonian action \eqref{eq:Lphi} is understood as the effective action in which tree level and loop level effects have already been integrated out, and light loops are yet to be integrated out. Cutoff schemes are known to typically violate decoupling and are sensitive to field redefinitions \cite{Burgess:1992gx,Kaplan:2005es,deRham:2014wfa,Gripaios:2015qya} and so these loops are computed in dimensional regularization.  }.\\

Although the Lagrangian terms may have multiple suppression scales, we stress that the theory still only has a single strong coupling scale, determined by the breakdown of perturbative unitarity. Generically, we expect the higher derivative operators to break perturbative unitarity at energies $s \sim \text{Min} \left\{ \Lambda_{\phi}^2 , \Lambda_A^2 \right\}$.

\subsection{Proca EFT}

In what follows, as a specific example, we consider for simplicity the (technically natural) hierarchy $\Lambda_\phi^3=\Lambda_A^2 m$. One motivation behind this choice is the existence of a  natural decoupling limit defined by taking the massless limit $m\to 0$, while keeping the scale $\Lambda_{\phi}$ fixed so that $\Lambda_A \rightarrow \infty$. In this limit the scalar decouples from the vector, i.e. we obtain a free Maxwell theory and an interacting Goldstone mode $\phi$. This decoupling limit describes at the level of the Lagrangian, the essence of the Goldstone equivalence theorem, that in the high energy limit $ m\ll E \ll \Lambda_A$,  the scattering amplitudes are dominated by the Goldstone/\stu mode. While this is an interesting motivation, we note however that there is a priori no need to set $\Lambda_\phi^3=\Lambda_A^2 m$. and in order to apply the positivity bounds however we must maintain $m \neq 0$ which is required so that the Froissart-Martin bound (which determines the number of subtractions in the dispersion relation) applies. For this reason we need to work beyond the decoupling limit. \\

Since we are interested in applying the tree-level $2-2$ amplitude positivity bounds, it is sufficient to focus on the following contributions to the Proca EFT\footnote{
Note that to compare with Ref.~\cite{Bonifacio:2016wcb} one may replace our coefficients with $a_1 = \lambda_3, a_0 = - \lambda_6 m^2/\Lambda_\phi^2,  a_3 = \lambda_7, a_4 = \lambda_5 - 2 \lambda_6 - \lambda_7 , a_5 = -\lambda_4 + \lambda_5/2 - \lambda_6$ and take $\Lambda_{\phi}^3 = m^2 \mpl $ and $\Lambda_A^2 = m \mpl  $. Ref.~\cite{Bonifacio:2016wcb} does not explicitly contains a $A^4$ term but the term nonetheless arises by field redefinition. Note that in this approach, we have \emph{not} imposed the Galileon decoupling limit condition of \cite{Bonifacio:2016wcb}, as having operators at a scale $\Lambda_\phi$ with higher order equations of motion simply signals that the EFT can no longer be trusted at that scale, as already expected from the EFT approach. However after  completely removing any total derivatives or field redefinition redundancy we retain the same number of free parameters as in \cite{Bonifacio:2016wcb}.},
\begin{align}
\label{eqn:ProcaLeading}
g_*^2 \mathcal{L}_{\text{Proca}}  &\supset  - \frac{1}{4} F_{\mu}^{\nu} F^{\mu}_{\nu}
 - \frac{1}{2} \phi_\mu^2
 + \frac{ a_0  }{\Lambda_\phi^4  }  \phi_\mu^4
+  \frac{a_1}{ \Lambda_\phi^3 } \partial_\mu  \phi_\nu  \phi^\mu \phi^\nu
 \\
&  + \frac{1}{\Lambda_\phi^6} \left(
a_3   ( \phi_\mu \partial^\mu \phi_\nu  )^2
+ a_4 (  \partial_\mu \phi_\nu  \phi^\nu   )^2
  + a_5  \phi_\mu^2 \partial_\alpha  \phi_\beta  \partial^\beta \phi^\alpha
\right)
  \nonumber \\
  &+  \frac{1}{\Lambda_A^4} \left( c_1  F^\mu_{\, \nu} F^\nu_{\, \rho} F^\rho_{\, \sigma} F^\sigma_{\, \mu}  +  c_2  (F_{\, \mu}^{\nu} F^{\mu}_{\, \nu}   )^2   \right)
\nonumber \\
&+ \frac{m^4}{\Lambda_{\phi}^6}  \left(  C_1 \phi_\mu \phi^\nu F^{\alpha \mu} F_{\alpha \nu}  + C_2 \phi_\mu^2 F_{\alpha\beta}^2   \right)\,,\nn
\end{align}
where we have used extensively integration by parts and field redefinition to eliminate redundant terms.
As we have discussed above, we expect $c_{1,2}/\Lambda_{A}^4$ to be at least of order $m^4/{\Lambda_{
\phi}^8}$ but could be much larger. With the choice $\Lambda_\phi^3=\Lambda_A^2 m$ this will be of order $m^2/{\Lambda_{\phi}}^6$ and will the contribute terms of the same order as $a_{3,4,5}$ and $C_{1,2}$. We can further remove the cubic term with coefficient $a_1$ by performing the non-linear local field redefinition $\phi\to \phi-a_1/(4\Lambda_\phi^3)\phi_\mu^2$ and absorb the quartic contribution into a redefinition of $a_4$. Without loss of generality we can therefore ignore the cubic operator going as $a_1$ in the rest of the discussion.\\

For explicit calculations of scattering amplitudes, it is convenient to choose unitary gauge $\phi = 0$, so the EFT becomes
\begin{align}
\label{Lprouni}
g_*^2 \mathcal{L}_{\text{Proca}}^{\text{unitary}} &\supset - \frac{1}{4}  F_{\mu}^{\nu} F^{\mu}_{\nu} - \frac{1}{2} m^2 A_\mu A^\mu
+ \frac{ m^4 a_0}{ \Lambda_{\phi}^4} \left( A_\mu A^\mu\right)^2
  \\
& + \frac{m^4}{\Lambda_{\phi}^6} \left(    a_3  A_\mu A_\nu  \partial^\mu A_\rho \partial^\nu A^\rho + a_4 A_\mu A_\nu  \partial_\rho A^\mu \partial^\rho A^\nu + a_5   A_\mu A^\mu \partial_\alpha A_\beta \partial^{\beta} A^{\alpha}     \right)
\nonumber \\
 &+  \frac{1}{\Lambda_A^4} \left( c_1  F^\mu_\nu F^\nu_\rho F^\rho_\sigma F^\sigma_\mu  +  c_2 (F\mn^2   )^2  \right)
+ \frac{m^4}{\Lambda_{\phi}^6} \left(  C_1 A_{\mu}A^\nu F^{\alpha \mu} F_{\alpha \nu}    + C_2 F\mn^2\,   A_\alpha A^\alpha  \right)\nonumber\,.
\end{align}
Here and in the following, by using $\supset$, we mean only to include terms that will contribute to the leading order in the amplitude (that is, power counting at the level of amplitudes). Note that the suppression scales in unitary gauge looks far from intuitive, which is the reason that we should always power-count in the \stu formalism and then take unitary gauge. In the following, we will constrain the parameter space of this theory using the positivity bounds.

%%%
\subsection{Scattering Amplitudes}
To compute the amplitudes, we use the polarization vectors in the transversity basis as given in \cite{deRham:2017zjm}
\bal
\ep^\mu_{\ti=\pm1} &= \frac{i}{\sqrt{2}m}( p, E \sin \theta  \pm im\cos\theta, 0, E\cos\theta \mp im\sin\theta ) ,
\\
\ep^\mu_{\ti=0} &= (0,0,1,0)  ,
\eal
where $E$ and $p$ are the energy and the absolute value of the momentum in the center of mass frame.
To leading order at tree-level, the only independent elastic amplitudes in the transversity basis are (up to an overall factor of $1/g_*^2$)
\ba
&& \T^+_{0000}=2s^2\tilde s^2\(24 \frac{m^4}{\Lambda_\phi^4} a_0 -8\frac{m^6}{\Lambda_\phi^6}\(a_4+C_1+2C_2\)
+8\frac{m^2\(6 m^4+x \)}{\Lambda_\phi^6}\(\tilde c_1+2\tilde c_2\)\) \, ,  \\
&& \T^+_{-11-11}=2s^2\tilde s^2\Bigg[\frac{x-4m^2(t-4m^2)}{\Lambda_\phi^4}\(a_0-\frac12\frac{m^2}{\Lambda_\phi^2}(a_4-4(\tilde c_1+2\tilde c_2)+C_1)\) \nn\\
&& +\frac{3}{8}\frac{y}{\Lambda_\phi^6}\(a_3+a_4-2a_5\)-\frac{m^2 s u}{\Lambda_\phi^6}
\(\frac32 a_3-a_4+a_5+\frac 32 C_1+2C_2\)\Bigg] \, \, ,   \\
&& \T^+_{0101}=\frac{m^2s^2 \tilde s (st-4m^2u)}{\Lambda_\phi^4}\,
\Bigg[4a_0-\frac12\frac{u}{\Lambda_\phi^2}(a_3+C_1)+2 \frac{s-t}{\Lambda_\phi^2}\tilde c_1\\
&&+\frac{t}{\Lambda_\phi^2}(-a_4+a_5)-4\frac{2t-4m^2}{\Lambda_\phi^2}\tilde c_2+2\frac{t-4m^2}{\Lambda_\phi^2}C_2\Bigg]
+\frac{m^2s^2\tilde s^3(s-u)}{2\Lambda_\phi^6}(a_3+4\tilde c_1+C_1) \, , \nn\\
&& \T^+_{1111}=\frac{2s^2}{\Lambda_\phi^4}\left[\tilde s^2(t^2+t \tilde s+\tilde s^2)+4m^2 s (8t^2+8t \tilde s+\tilde s^2)\right]\(a_0+\frac{2m^2}{\Lambda_\phi^2}(\tilde c_1+2\tilde c_2-C_2)\) \nn \\
&& +\frac{s^2 \tilde s}{4 \Lambda_\phi^6}\left[\tilde s^2(4m^2 s -3 tu)+16m^2t(t+\tilde s)(3s-4m^2)\right]\(a_3+a_4-2a_5\)\,,
\ea
where we write $\tilde s=s-4m^2$ and $\tilde c_{1,2} =c_{1,2} \Lambda_{\phi}^6/(m^2 \Lambda_A^4)$,
and keep otherwise a similar notation as in \cite{deRham:2017avq,deRham:2017imi} and denote the Lorentz crossing-symmetric invariants as $x=-(s t+s u+u t)$ and $y=-s t u$.  \\

Even though we will apply most of our bounds in the definite transversity basis, it is useful to define quantities which are valid for generic states,
\ba
 \alpha_{\pm} \equiv \frac{1}{\sqrt{2}} \left( \alpha_{-1} \pm \alpha_{+1} \right) \,,\qquad \beta_{\pm } \equiv \frac{1}{\sqrt{2}}  \left( \beta_{-1} \pm \beta_{+1} \right)\,.
\ea
where  $\alpha_{\pm 1, 0}, \beta_{\pm 1, 0}$ designate the projection along the definite polarization vectors in the transversity basis.
In what follows we first apply the positivity bounds in the forward limit ($t=0$) and recover the results presented in \cite{Bonifacio:2016wcb}. We then consider definite transversity bounds beyond the forward limit.\vspace{0.3cm}

%%%
\paragraph{Forward Limit:}
The contributions of the previous operators to the  forward  limit bound is\footnote{For the forward limit, we may extend the bounds to generic indefinite transversity (or helicity) states.}
\ba
\label{eq:fab}
f_{\alpha\beta}\Big|_{t=0} &=& \frac{8}{\Lambda_\phi^4}\(a_0
-\frac 12 \frac{m^2}{\Lambda_\phi^2}\(a_4+C_1\)\)| \alpha_+ |^2 | \beta_+ |^2\\
&+&\frac{4m^2}{\Lambda_\phi^6}\(a_3-2a_4+2a_5+C_1+4C_2\)
\(\text{Re}[\alpha_0^* \alpha_+]\text{Re}[\beta_0^* \beta_+]
-\text{Re}[\alpha_-^* \alpha_+]\text{Re}[\beta_-^* \beta_+]\)\nn\\
&+&\frac{2m^2}{\Lambda_\phi^6}\(a_3+C_1\)\( | \alpha_+|^2 | \beta |^2 + |\alpha |^2 | \beta_+ |^2  \) \nn\\
&+&\frac{8m^2}{\Lambda_\phi^6}\tilde c_1\(| \alpha_0|^2 + | \alpha_- |^2 \) \( | \beta_0|^2 + | \beta_- |^2  \)\nn\\
&+&\frac{8m^2}{\Lambda_\phi^6}(\tilde c_1+4\tilde c_2)
\(\left| \alpha_0 \beta_0 - \alpha_- \beta_-    \right|^2
-2{\rm Im}[\alpha_0^*\alpha_-]{\rm Im}[\beta_0^*\beta_-]\)\,.\nn
\ea
This can be simplified by exploiting the normalization $| \alpha |^2 = | \beta |^2 =1$ as we shall do below. However first we notice that so long as $a_0>0$, we need not worry about the contributions from the second line. Indeed the second line can only contribute if $\alpha_+ \beta_+\ne 0$, in which case the term proportional to $a_0$ dominates anyways (as we shall see below in section~\ref{sec:galileon}, the situation is quite different if the scaling of the operator $A_\mu^4$ is taken differently). \\

With $a_0>0$ as our first requirement, the rest of the bounds can be determined by assuming $\beta_+=0$ without loss of generality. The indefinite positivity bounds then require
\ba
&&\(a_3+C_1\) | \alpha_+|^2 +4 \tilde c_1\(1-|\alpha_+|^2 \)\\
&+&4(\tilde c_1+4 \tilde c_2)
\(\left| \alpha_0 \beta_0 - \alpha_- \beta_-    \right|^2
-2{\rm Im}[\alpha_0^*\alpha_-]{\rm Im}[\beta_0^*\beta_-]\)>0 \,,\nn
\ea
for all choices of normalized states with $\beta_+=0$.

\begin{itemize}
\item Now if $\Lambda_A^2 \gg \Lambda_{\phi}^3/m$, this implies $\tilde c_{1,2} \ll 1$ and the above bounds will simply be satisfied by
\be
a_0 > 0 \,, \qquad \quad {\rm and }\quad a_3+C_1>0 \,,  \, \quad \text{ for } \Lambda_A^2 \gg \Lambda_{\phi}^3/m\,.
\ee
\item On the other hand if $\Lambda_A^2 = \Lambda_{\phi}^3/m$ then choosing say $\alpha_0=\beta_-=0$, we see immediately that the first line should be positive, leading to the two conditions $a_3+C_1>0$ and $c_1>0$. Finally by spanning over the possible states, we see that the only last condition is $c_1+8c_2>0$, so in summary, the positivity bounds set the requirements
\ba
\label{eq:bound1Proca}
 a_0 > 0 \,, \qquad c_1>0\,, \qquad c_1+2c_2 > 0 \,, \quad {\rm and }\quad a_3+C_1>0 \,, \quad \text{ for } \Lambda_A^2 = \Lambda_{\phi}^3/m\,. \qquad
\ea
\end{itemize}
At this level we can also notice that truncating the Proca EFT at that stage would lead to bounds that would {\it seem} to be violated at that order, for instance for  states with $\alpha_{\pm 1}= \pm 1/\sqrt{2}$, $\alpha_0=0$ and $\beta_{\pm 1}=1/\sqrt{2}$, $\beta_0=0$ we would have $f_{\alpha \beta}=0$. In reality this simply expresses the need for  higher order operators to enter the EFT and contribute at order $m^4/\Lambda_\phi^8$ in $f$.

\vspace{0.3cm}

%%%
\paragraph{Probing beyond the forward limit:}
The leading contribution to the first $t$ derivative is,
\begin{equation}
\frac{\pd}{\pd t} f_{\ai\bi}\Big|_{t=0}  =\frac34 \frac{a_3+a_4-2a_5}{\Lambda_\phi^6}  \; | \alpha_+ |^2 | \beta_+ |^2\,.
\label{eqn:ProcadtLeading}
\end{equation}
Remembering the general form of the first $t$ derivative bound
\be
\frac{\pd}{\pd t} f_{\ti_1\ti_2}  + \frac{N_S+1}{2\mcl{M}^2} f_{\ti_1\ti_2}>0\,.
\ee
where here $N_S=2+8=10$ and since we are applying tree-level bounds then ${\cal M}^2= \Lambda_{\rm th}^2-2m^2 \sim \Lambda_{\rm th}^2$ where $\Lambda_{\rm th}$ is the mass of the next state the lightest state not included in the EFT, i.e. the cutoff. Since in a weakly coupled UV completion we expect $\Lambda_{\phi} \sim \Lambda_{\rm th}$, i.e. the interactions of the Goldstone arise from integrating out the massive modes, we may choose to define the Lagrangian parameters so that $\Lambda_{\phi} =\Lambda_{\rm th}$. Then with only the further assumption of the hierarchy $m \ll \Lambda_{\phi}$ the first $t$ derivative positivity bound amounts to
\be
3(a_3+a_4-2a_5)+ 112 a_0 \gtrapprox 0 \, .
\ee
Note that as stated this is a linear condition on dimensional coefficients which are expected to be of order unity in a Wilsonian sense. Crucially this is quite independent of the forward limit bounds and so provides new information on the parameter space and constrains the operators of the form $A^2 (\p A)^2$ in a way which would not have been possible without going beyond the forward limit. This demonstrates the usefulness of non-forward limit positivity bounds.\\

These bounds are the ones derived for a generic Proca EFT with the scale counting as explained in section~\ref{sec:scales}. Note that this scaling differs from that we would encounter in a ``charged Galileon" effective theory where a Galileon shift symmetry is imposed for the helicity-0 mode, symmetry which is then only softly broken by other operators. In such an EFT, operators that break the Galileon invariance (for instance operators of the form $(\p \phi)^4$) are required to be additionally suppressed by powers of $m^2/\Lambda_\phi^2$. Such a tuning is expected to be stable and hence self-consistent \cite{Luty:2003vm,Nicolis:2004qq,deRham:2012ew}.

%%%%%%%%%%%%%%%%%%%%%%%%%%%%
\section{Charged Galileon}
\label{sec:galileon}
%%%%%%%%%%%%%%%%%%%%%%%%%%%%

If we consider a theory with a global Galileon symmetry
\ba
 \phi ( x ) \to  \phi (x) + c + b_\mu x^\mu,
\ea
that is only softly broken, all operators that break this Galileon symmetry are then naturally suppressed by the symmetry breaking parameter. In practise, when considering a generic scalar field EFT with an unbroken shift symmetry $\phi\to \phi+c$, the soft breaking of an additional Galileon symmetry suppresses  all operators of the form $(\partial \phi)^n$. This enhanced soft behaviour is related to $\phi$'s interpretation as a Goldstone for spontaneously broken diffeomorphisms in massive theories of gravity (see for example \cite{deRham:2014zqa,deRham:2016nuf}). \\

Now consider a nonlinearly realized gauge group, under which the scalar field transforms as in
\ba
A_\mu \to A_\mu + \partial_\mu \epsilon, \;\;\;\; \phi \to \phi - m \epsilon .
\label{eqn:ProcaStuckelberg}
\ea
Generically, the leading $(\partial \phi)^4/ \Lambda^4$ is gauged into $(D \phi)^4/\Lambda^4 \supset m^4 A^4 / \Lambda^4$, which was the operator that dominated the positivity bounds in Section~\ref{sec:proca} (term governed by the coefficient $a_0$). However, if the original scalar enjoys a Galileon shift symmetry, which forbids the coefficient of $(\partial \phi)^4/\Lambda^4$, then on introducing the gauge field we find an EFT of the form \eqref{eqn:ProcaLeading} but with now $a_0 \sim m^2/\Lambda_{\phi}^2$ in order to respect the soft Galileon amplitude behaviour. This particular tuning, $a_0 \sim m^2/ \Lambda^2$, is stable against loop corrections due to the Galileon symmetry.

%%%
\paragraph{Decoupling Limit:}
As we have seen in the Proca theory, for an EFT with a nonlinearly realized gauge symmetry, employing the \stu field was useful in correctly assessing the suppression scales of the operators.  Before turning to the positivity bounds for this charged Galileon EFT, we wish to make a brief digression on the construction of such EFTs directly in unitary gauge.\\

In a standard Wilsonian EFT picture, all operators that satisfy a given symmetry ought to be introduced at the cutoff scale of the theory, and there is no requirement of `ghost-freedom' in the sense that the equations of motion are second order at that scale. Higher derivative operators are allowed provided they are suppressed by the cutoff, so that the associated ghost is at or above the scalar of the cutoff, meaning it does not exist.  \\

However in some more recent constructions (in particular those related to the existence of a Vainshtein mechanism or a strong coupling), attempts to build theories that could {\it in principle} be trusted beyond their naive strong coupling scale by virtue of the Vainshtein mechanism have been made \cite{Heisenberg:2014rta}. The Galileon EFT for which no \Ost mode would be present at the strong coupling scale therefore takes on a special role among such EFT.   For example, the associated charged Galileon theory with second order equations of motion in unitary gauge would be constructed as
\begin{align}
\label{eqn:ProcaLwithDL}
\mathcal{L} &= - \frac{1}{4} F\mn^2  -  \frac{m^2}{2} A_\mu A^\mu +  \frac{ g_3  m}{3! \mpl } A^\mu A^\nu \partial_\mu A_\nu  + \frac{ g_4 }{4! \mpl ^2}  \delta^{\mu_1 \mu_2}_{\nu_1 \nu_2} A^\mu A_\mu \partial_{\mu_1} A^{\nu_1} \partial_{\mu_2} A^{\nu_2} \\
 &+ \frac{\alpha_1}{\mpl ^2} \delta^{\mu_1 \mu_2}_{\nu_1 \nu_2}   A_\alpha \partial_{\mu_1} A^\alpha A_{\mu_2} \partial^{\nu_1} A^{\nu_2} + \frac{\alpha_2}{\mpl ^2} \delta^{\mu_1 \mu_2}_{\nu_1 \nu_2} A_\mu A^\mu \partial_{\mu_1} A_{\mu_2} \partial^{\nu_1} A^{\nu_2}
 \nn\\
&  + \frac{\alpha_3}{\mpl ^2}  \delta^{\mu_1 \mu_2 \mu_3}_{\nu_1 \nu_2 \nu_3} A_{\mu_1} A^{\nu_1} \partial_{\mu_2} A^{\nu_2} \partial^{\nu_3} A_{\mu_3} +  \frac{1}{\Lambda_A^4} \left( c_1  F^\mu_{\, \nu} F^\nu_{\, \rho} F^\rho_{\, \sigma} F^\sigma_{\, \mu}  +  c_2  (F_{\, \mu}^{\nu} F^{\mu}_{\, \nu}   )^2   \right) \,,\nn
\end{align}
which corresponds to the choice of $\Lambda_{\phi}^3 = m^2 \mpl $.
After appropriately introduce the \stu field, $A_\mu \to A_\mu + \partial_\mu \phi / m$,
this choice of scales ensures that the Lagrangian reduces to a Galileon theory (in $\phi$) in the limit where $m\to 0$, $\mpl\to \infty$ while keeping the scale $\Lambda_\phi=(m^2 \mpl)^{1/3}$ fixed.\\

As already mentioned, from a standard  Wilsonian EFT point of view we would expect higher derivative operators to be generating when integrating out degrees of freedom from the UV theory. The associated would-be ghost then has a mass at the scale of the removed heavy particles, i.e. the cutoff of the EFT, and therefore does not exist. From a strict bottom-up perspective, one should simply write down all of the available local operators and any ghost from a higher derivative operator that is introduced should simply be interpreted as imposing an upper bound for the cutoff of the theory. Nevertheless, focusing on those theories that do not involve higher derivatives has its own merit in some physical context, particularly when one is interested for phenomenological reasons in a re-organization of the effective field theory. Interestingly, within the context of this study, where we are only interested in the leading order in 2-to-2 tree-level scattering, the two approaches are equivalent. This is because, starting from any arbitrary local Lagrangian in terms like $\partial A^3$ and $\partial^2 A^4$, one can always perform a field redefinition to bring it to the form \eqref{eqn:ProcaLwithDL}  (to this order in power counting at the level of amplitudes and up to total derivatives). Of course, going beyond leading order, this will no longer remain the case (although the order at which the first higher-derivative enter can be high \cite{Solomon:2017nlh}). \\

Therefore, for a charged Galileon, we may continue using the Proca theory defined in Eqn.~\eqref{Lprouni}, with the vital difference that for the charged Galileon the term governed by $a_0$ has an extra $m^2/\Lambda_\phi^2$ suppression. In practise we simply use the operators present in \eqref{Lprouni}, with $a_0\to a_0m^2/\Lambda_\phi^2$, so that the quartic operator $A^4$ is now governed by the coefficient $a_0 m^6/\Lambda_\phi^6$. The leading terms of most of the scattering amplitudes are hence no longer dominated by this operator which significantly changes the bounds imposed both in the forward limit and beyond.

%%%
\paragraph{Forward limit:}
The forward limit bounds can be read of from the expression for $f_{\alpha \beta}$ given in \eqref{eq:fab} with again $a_0\to a_0m^2/\Lambda_\phi^2$. This implies that $a_0>0$ is not necessarily sufficient to guarantee the first two lines of \eqref{eq:fab} to be positive.

%%%
\paragraph{Necessary conditions:}
For simplicity we re-write the quantity $f_{\alpha\beta}$ in terms of five relevant parameters
\ba
\label{eq:fab2}
f_{\alpha\beta}\Big|_{t=0} &=&
\frac{8m^2}{\Lambda_\phi^6}\Bigg[
\mu_1| \alpha_+ |^2 | \beta_+ |^2
+\mu_2\( | \alpha_+|^2 +  | \beta_+ |^2  \) \\
&+&\mu_3\( \text{Re}[\alpha_-^* \alpha_+]\text{Re}[\beta_-^* \beta_+] - \text{Re}[\alpha_0^* \alpha_+]\text{Re}[\beta_0^* \beta_+] \)\nn \\
&+&\mu_4\(| \alpha_0|^2 + | \alpha_- |^2 \) \( | \beta_0|^2 + | \beta_- |^2  \)\nn\\
&+&(\mu_5-\mu_4)
\(\left| \alpha_0 \beta_0 - \alpha_- \beta_-    \right|^2
-2{\rm Im}[\alpha_0^*\alpha_-]{\rm Im}[\beta_0^*\beta_-]\)\Bigg]\,.\nn
\ea
with
\ba
& \mu_1=a_0-\frac 12 \(a_4+C_1\)\,,\quad
\mu_2=\frac14\(a_3+ C_1\)\,,\quad
\mu_3= a_4 - a_5 - 2C_2 - \frac 12 (a_3+C_1)   \label{eqn:Proca_mu}\\
& \mu_4=\tilde c_1\qquad {\rm and }\qquad \mu_5=2(\tilde c_1+2 \tilde c_2)\,.\nn
\ea
In this form, we quickly see that the positivity bound requires
$\mu_1 + 2\mu_2 >0$, $\mu_4>0$ and $\mu_5>0$. In addition to these four bounds (which had an equivalent in the Proca EFT \eqref{eq:bound1Proca}), there are now new bounds on $\mu_3$, such as $|\mu_3|<X$, where the precise expression for $X$ is non-trivial to compute explicitly, but $X$ should be at least
\ba
\label{eq:eqX}
X\le {\rm Min}\left[\mu_1+4\mu_2+\mu_5,\frac 12 (\mu_1+6\mu_2+4\mu_5)\right].
\ea
However, this is not necessarily the strictest requirement on $\mu_3$. Besides resorting to a numerical minimization of $f_{\alpha \beta}$, one can derive some sufficient conditions for positivity provided in appendix~\ref{App:sufficientCond}.\\

Altogether, we can express these conditions as,
\begin{equation}
a_0 + \frac{1}{2} a_3 - \frac{1}{2} a_4 > 0  ,  \;\; \tilde c_1 > \text{Min} [ 0 , - 2 \tilde c_2 ]   , \quad {\rm and} \quad | \mu_3 | < X\,.
\end{equation}
The first bound corresponds to $SSSS$ scattering, and coincides with the result of \cite{Bonifacio:2016wcb}.
The final bounds depends in general on a non-trivial function of the EFT parameters, $X$,  which can be determined numerically. Note that $X$ always lies in the range \eqref{eq:eqX}. \vspace{0.5cm}

\begin{figure}
\centering
\includegraphics[width=0.75\textwidth]{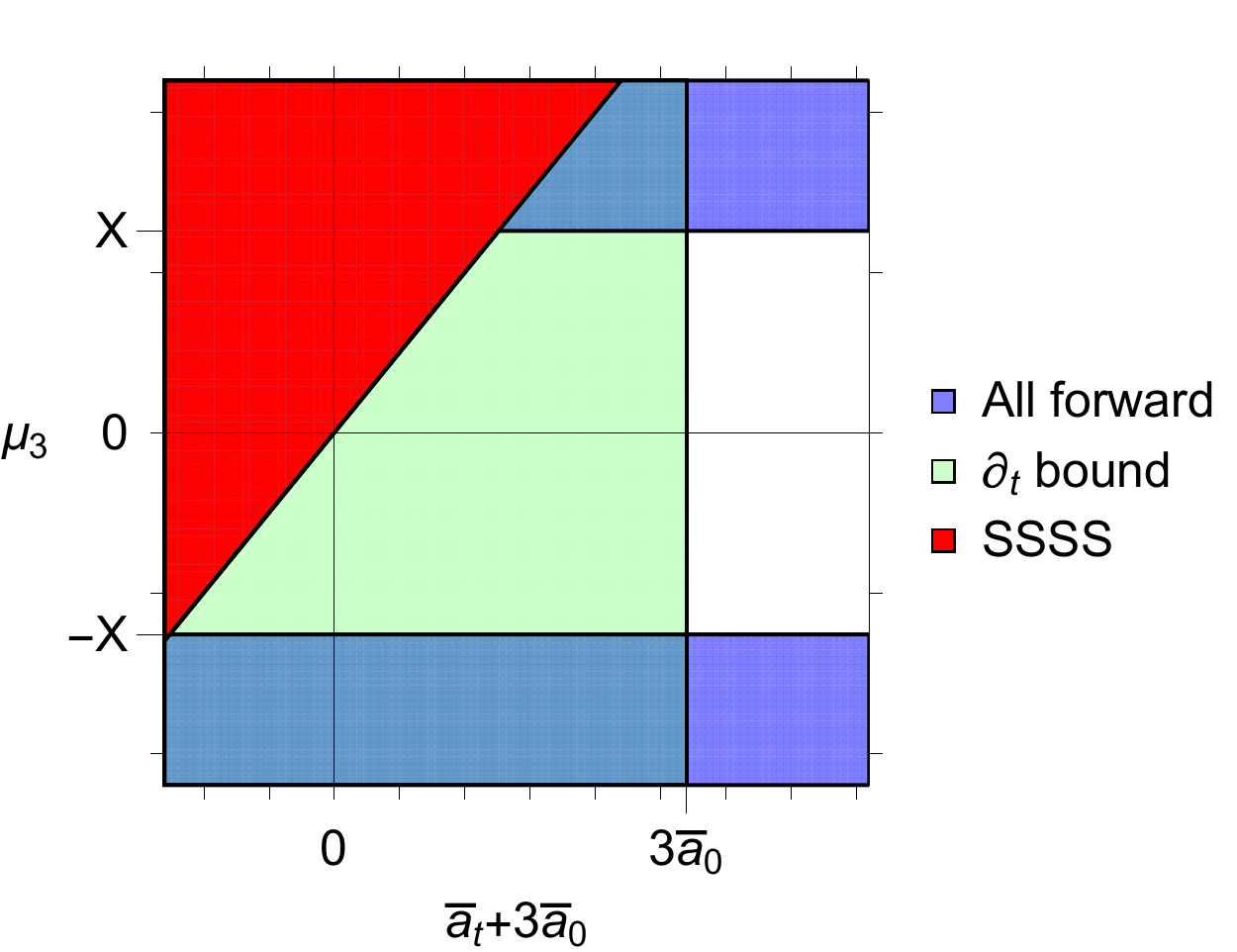}
\caption{ Parameter space for charged Galileon theory constrained by analyticity.
A suppressed $a_0 \sim m^2/\Lambda_{\phi}^2$ is required to recover a Galileon symmetry in the decoupling limit. The forward limit bound from scattering four scalar modes was previously found in \cite{Bonifacio:2016wcb}, which excluded the red region. By looking at scattering superpositions of modes, one use forward limit bounds to further restrict the parameter space to a semi-infinite strip (the white region plus the light green region) in the parameter space spanned by $( \mu_3 ,  \bar a_t + 3\bar a_0  )$. Utilizing the first $\partial_t$ bound (i.e., going away from the limited forward limit formalism), we reduce this semi-infinite strip even further, potentially greatly, depending on the sign of $3 \bar a_0 = 3 a_0 - a_5 - C_1 - 4 C_2$. In terms of the EFT coefficients in \eqref{eqn:ProcaLeading}, we have $2 \bar a_t = a_3 + a_4 - 2 a_5,\; \mu_3 = a_4 - a_5 - a_3/2 - C_1/2 - 2C_2 $. The $SSSS$ bound corresponds to $\bar{a}_t + 3 \bar{a}_0 - 2 \mu_3 > 0$.
}
\label{fig:Proca}
\end{figure}

%%%
\paragraph{First $t$ derivative:}
The leading contribution to $\partial f_{\ti_1\ti_2}/\pd t$ \eqref{eqn:ProcadtLeading} is not affected by the ordering of $a_0$. However, given the current scaling of  $a_0$ then the contribution going as  $f_{\alpha \beta}$ in \eqref{eq:2ndbound} is suppressed by $m^2/\mathcal{M}^2\ll 1$ as compared with the term coming from $\p_t f_{\alpha\beta}$ and so the positivity bounds require the quantity $\p_t f_{\alpha\beta}$ \eqref{eqn:ProcadtLeading} to be positive by itself,
hence leading to
\ba
\bar a_t =  \frac{1}{2} a_3 + \frac{1}{2} a_4 - a_5   >  0\,.
\ea
Again, going away from the forward limit in this way has given us a new constraint which could not have been derived from forward limit considerations alone.
A selection of these bounds is plotted in Figure \ref{fig:Proca}, demonstrating that this $t$ derivative constraint can be orthogonal to the previous forward limit constants.

%%%%%%%%%%%%%%%%%%%%%%%%%%%
%%%%%%%%%%%%%%%%%%%%%%%%%%%

\section{Simplest Vector Galileon}
\label{sec:SimplestVec}

To illustrate the power of the positivity bounds in constraining EFTs which would otherwise seem valid, consider the following tuning of \eqref{eqn:ProcaLeading}: Set to zero all of the Wilson coefficients\footnote{From a standard EFT viewpoint, we would expect the Wilsonian coefficients to run under quantum corrections, so this tuning is not radiatively stable, but this is not the concern of this section.} at this order, \emph{except} $a_1$ and $a_5$.  This is equivalent, up to a field redefinition and boundary terms, to the following unitary gauge action,
\begin{equation}
\mathcal{L} = - \frac{1}{4} F^\nu_\mu F^\mu_\nu - \frac{1}{2} m^2 A_\mu^2 + \frac{g_3 m^3}{\Lambda^3} A_\mu A^\mu \partial^\nu A_\nu + \frac{g_4 m^4}{\Lambda^6} A_\mu A^\mu \left(  (\partial_\nu A^\nu )^2 - \partial_\alpha A^\beta \partial_\beta A^\alpha    \right)\,.
\label{eqn:simplest_vector}
\end{equation}
Due to its remarkable simplicity and similarity to the scalar Galileon, the above action is a particularly simple realization of Proca theory which has provided a useful testing ground for many ideas in recent literature \cite{Tasinato:2014eka,Gripaios:2004ms,Heisenberg:2014rta}. For example, its covariant coupling to gravity has been considered and remarkably it was shown to admit black hole vector hair \cite{Heisenberg:2017hwb, Heisenberg:2017xda,Chagoya:2016aar}. Interesting numerical simulations of neutron stars have also been considered \cite{Kase:2017egk} in this model. Since such a model has proven to be of great interest phenomenologically, we may wonder if and how it may be realized within a standard Wilsonian completion. \\

Positivity bounds show that a Wilsonian UV completion of the above vector Galileon would require,
\ba
&& \Lambda^4 f_{-1 1}  = \frac{2 m^2}{\Lambda^2}  ( -3 g_3^2  - g_4 )
%+ \frac{3 t}{\Lambda^6} ( g_3^2 + g_4)
 > 0 \\
&& \Lambda^4 f_{11} =  \frac{2 m^2}{\Lambda^6} ( g_3^2 + g_4 )
%+ \frac{3 t}{\Lambda^6} ( g_3^2 + g_4 )
> 0
\ea
or put another way,
\begin{equation}
3 g_3^2 < - g_4 < g_3^2\,.
\end{equation}
This is clearly not possible unless $g_3 =  g_4 = 0$ at this order, and subleading terms in the EFT come in to satisfy these bounds (or the theory is free to all order).\\

So the theory \eqref{eqn:simplest_vector} cannot admit a Lorentz invariant UV completion unless additional operators are included at the same order. An example of such an operator, which still preserves the `ghost-freedom' of the original theory, is,
\ba
\mathcal{L} &=& - \frac{1}{4} F^\nu_\mu F^\mu_\nu - \frac{1}{2} m^2 A_\mu^2 + \frac{g_3 m^3}{\Lambda^3} A_\mu A^\mu \partial^\nu A_\nu + \frac{g_4 m^4}{\Lambda^6} A_\mu A^\mu \left(  (\partial_\nu A^\nu )^2 - \partial_\alpha A^\beta \partial_\beta A^\alpha    \right)  \\
&&+ \frac{C_1 m^2}{\Lambda^6} ( A^\mu F_{\alpha \mu} )^2 \,.\nn
\ea
The positivity bounds are then,
\ba
&& \Lambda^4 f_{-1 1}  =  \frac{2 m^2}{\Lambda^2}  ( C_1 - 3 g_3^2  - g_4 )
{>} 0 \\
&& \Lambda^4 f_{01} = \frac{m^2}{\Lambda^6} C_1 {>} 0  \\
&& \Lambda^4 f_{11} =  \frac{2 m^2}{\Lambda^6} ( g_3^2 + g_4 )
 {>} 0 .
\ea
This opens up an allowed window of parameter space,
\begin{equation}
3 g_3^2 - C_1 < -g_4 < g_3^2  \;\;\;\; \implies \;\;\;\; C_1 {>} 2 g_3^2 >  -2 g_4 .
\end{equation}
Within this framework, we therefore see that  in order for the Proca model presented in \eqref{eqn:simplest_vector} to admit a standard Wilsonian UV completion, other operators are forced upon us, with coefficients which are of the same order as $g_3$ or $g_4$. \\

An interesting question is how the phenomenology indicated in \cite{Heisenberg:2017hwb, Heisenberg:2017xda,Chagoya:2016aar,Kase:2017egk} is affected by the existence of these types of additional operators.
Related to this is the question of the vacuum. Technically speaking, the positivity bounds make sense when considering a trivial vacuum solution on Minkowski.
Yet the Proca model in \eqref{eqn:simplest_vector} has many cosmological and astrophysical applications that naturally take place on non-trivial backgrounds (and non-trivial field configurations). However so long as the theory also admits a Minkowski vacuum (and so long as one can extrapolate between these vacua while remaining within the regime of validity of the LEEFT), the bounds derived here are applicable independently of the background on which one decides to use the theory for phenomenological purposes.

%%%%%%%%%%%%%%%%%%%%%%%%%%%%
\section{$\Lambda_5$ Massive Gravity}
\label{sec:L5}
%%%%%%%%%%%%%%%%%%%%%%%%%%%%

In a generic theory of Lorentz invariant massive gravity, the first natural scale that appears is the $\Lambda_5$ scale, where $\Lambda_5^5 = \mpl m^4$. For instance, if we take the Einstein-Hilbert kinetic term, and simply add the Fierz-Pauli mass term for the metric perturbation $h_{\mu\nu}=g_{\mu\nu}-\eta_{\mu\nu}$, it is straightforward to show that such a theory breaks perturbative unitarity at the scale $\Lambda_5$ \cite{ArkaniHamed:2002sp}. In fact, in the absence of any specially tuned structure, this is the highest scale to which such an EFT can be considered. As is now well known, there is a special structure which raises the scale to $\Lambda_3$ where $\Lambda_3^3=m^2\mpl$ \cite{deRham:2010gu,deRham:2010ik,deRham:2010kj}. Remarkably we shall find below that the positivity bounds themselves will impose precisely this specially tuned structure that raises the $\Lambda_5$ scale to $\Lambda_3$, at least to quartic order. \\

In order to aid comparison with the $\Lambda_3$ theory to be discussed later, it is helpful to introduce the \stu fields
\be
\phi^a = x^a - \frac{V^a}{m \mpl}- \frac{\partial^a \pi}{m^2 \mpl}\,,
\ee
and define the tensor $K^{\mu}{}_{\nu}$ \cite{deRham:2010kj} via
\be
K^{\mu}{}_{\nu} = \delta^{\mu}{}_{\nu}-\sqrt{g^{\mu \omega} \partial_{\omega} \phi^a \partial_{\nu} \phi^b \eta_{ab}} \, .
\ee
In unitary gauge, $\phi^a = x^a$, then
\be
K^{\mu}{}_{\nu} = \frac{1}{2} h^{\mu}{}_{\nu} + {\cal O}(h^2) \, ,
\ee
implying that $K^{\mu}{}_{\nu} $ is just a particular choice of variables encoding the metric perturbations.
In the $\Lambda_5$ decoupling limit, defined by taking $\mpl \rightarrow \infty$ and $m \rightarrow 0$, keeping $\Lambda_5$ fixed, then
\be
K^{\mu}{}_{\nu} \rightarrow \frac{1}{m^2 \mpl} \partial^{\mu}\partial_{\mu} \pi = \frac{m^2}{\Lambda_5^5} \partial^{\mu}\partial_{\mu} \pi\,.
\ee
To understand why $\Lambda_5$ is the natural scale for massive gravity, it is sufficient to note that a generic mass potential in unitarity gauge of the form (index structure suppressed for interactions)
\be
{\cal L}_{\rm mass} \sim -\frac{1}{8}m^2 \mpl^2 \( h_{\mu \nu}^2-h^2 +  a h^3 + \dots \)
\ee
is re-expressible as
\be
{\cal L}_{\rm mass}   \sim -\frac{1}{2}m^2 \mpl^2 \( K^{\mu}_{\nu}K^{\nu}_{\mu}-K^2 +  a' K^3 + \dots \) \, .
\ee
Then focusing on the generic cubic interaction, in the $\Lambda_5$ decoupling limit we have
\be
 -\frac{a'}{2}m^2 \mpl^2 K^3 \rightarrow  -\frac{a'}{2} \Lambda_5^4 \( \frac{\partial \partial \pi}{\Lambda_5^3} \)^3 \, ,
\ee
meaning that unless we engineer the cubic interactions to vanish ($a'=0$), or be total derivatives in the decoupling limit (as in the case of the $\Lambda_3$ theory), we inevitably generate a dimension 5 operator at the scale $\Lambda_5$.
When interpreted as a Wilsonian EFT, we expect whatever physics resolves perturbative unitarity at the scale $\Lambda_5$ will generate an infinite number of other operators who shall naturally come in the form
\be
\Delta{\cal L} = \Lambda_5^4 L_0\( \frac{\Lambda_5^2}{m^2} K^{\mu}{}_{\nu}, \frac{\nabla_{\mu}}{\Lambda_5}, \frac{R^{\mu}{}_{\nu \rho \sigma}}{\Lambda_5^2} \) \, ,
\ee
where $L_0$ accounts for all scalar operators build out of its arguments, with order unity or smaller coefficients.
Thus the full form of the $\Lambda_5$ Wilsonian action is
\be
\label{eq:Lambda5_1}
S = \int \d^4 x \, \sqrt{-g} \left[  \frac{\mpl ^2}{2}  \left( R \left[ g \right] - m^2 \( K^{\mu}_{\nu}K^{\nu}_{\mu}-K^2\) \right) + \Lambda_5^4 L_0\( \frac{\Lambda_5^2}{m^2} K^{\mu}{}_{\nu}, \frac{\nabla_{\mu}}{\Lambda_5} , \frac{R^{\mu}{}_{\nu \rho \sigma}}{\Lambda_5^2}\)  \right] \, .
\ee
In particular we note that according to this counting, we can generate a $\Lambda_5^8 K^2/m^4 \sim (\Box \pi)^2/\Lambda_5^2 $ term which breaks the Fierz-Pauli tuned mass structure. However the associated Boulware-Deser ghost will have a mass at the scale $\Lambda_5$ and so from an EFT point of view such terms are allowed since the ghost cannot be excited in the regime of validity if the EFT. \\

A generalization of this to a `single scale- single coupling' theory (see \cite{deRham:2017xox} and references therein) by introducing a weak coupling parameter $g_*$ is
\be
\label{eq:Lambda5_2}
S = \frac{1}{g_*^2}\int \d^4 x \, \sqrt{-g} \left[  \frac{M^2}{2}  \left( R \left[ g \right] - m^2 \( K^{\mu}_{\nu}K^{\nu}_{\mu}-K^2\) \right) + \Lambda_5^4 L_0\( \frac{\Lambda_5^2}{m^2} K^{\mu}{}_{\nu}, \frac{\nabla_{\mu}}{\Lambda_5} , \frac{R^{\mu}{}_{\nu \rho \sigma}}{\Lambda_5^2}\)  \right] \, ,
\ee
where $M^2 = \mpl^2 g_*^2$ and $\Lambda_5$ is now $\Lambda_5=(m^4 M)^{1/5}$ is the cutoff of the effective theory. The virtue of introducing a weak coupling parameter $g_*\le 4 \pi$ is that loop corrections to the above effective action can be made arbitrarily small while leaving intact the form of the tree interactions, up to an overall normalization \cite{deRham:2017xox,Bellazzini:2017fep}. \\

For a weakly coupled UV completion for particles with spin $S  \ge 2$, the scattering amplitude must contain an infinite number of powers of $s$ since any finite order polynomial will scale at least as $s^S$, from the residue of the $t$-channel pole,  and so violates the fixed $ 0 \le t < m^2$ Froissart bound for $S \ge 2$. If the Froissart bound is violated by the tree amplitude, loop corrections become important at the energy for which this happens, which by definition contradicts our assumption that the theory has a weakly coupled UV completion. Since the amplitude will contain an infinite powers of $s$ (hence $t$ by crossing), as is well-known the UV completion will necessarily contain an infinite number of spins, as in the case of string theory. Indeed the weak coupling assumption seems always to lead to string-like behaviour at high energies for higher spin states \cite{Caron-Huot:2016icg}. \\

The generic form of the even part (which removes the $\sqrt{stu}$ branch cuts) of the tree-level 2-2 scattering amplitude for a generic polarization is then
\ba
&& \hspace{-1cm}{\cal T}_{\tau_1 \tau_2 \tau_3 \tau_4}(\theta)+ {\cal T}_{\tau_1 \tau_2 \tau_3 \tau_4}(-\theta) \sim  \\
&& \hspace{-1cm}\frac{g_*^2}{\Lambda_5^{10}} \Bigg[ \frac{  \sum_{a=0}^6 c^s_a t^a (m^2)^{6-a}}{m^2-s}+\frac{ \sum_{a=0}^6 c^t_a s^a (m^2)^{6-a}}{m^2-t}+\frac{  \sum_{a=0}^6 c^u_a s^a (m^2)^{6-a}}{m^2-u} + \hspace{-10pt}\sum_{a+b \le 5, a=0,b=0}^{a=5,b=5}\hspace{-10pt}  C_{ab} s^a t^b (m^2)^{5-a-b}\Bigg] \nn  \\
&& \hspace{-1.3cm}  +\frac{g_*^2}{\Lambda_5^{12}} \Bigg[ \frac{  \sum_{a=0}^7 d^s_a t^a (m^2)^{7-a}}{m^2-s}+\frac{ \sum_{a=0}^7 d^t_a s^a (m^2)^{7-a}}{m^2-t} +\frac{  \sum_{a=0}^7 d^u_a s^a (m^2)^{7-a}}{m^2-u} +\hspace{-10pt} \sum_{a+b \le 5, a=0,b=0}^{a=6,b=6} \hspace{-10pt} D_{ab} s^a t^b (m^2)^{6-a-b}\Bigg]  \nn
\\
&+& {\cal O}\(g_*^2/\Lambda_5^{14} \)+{\text{ kinematic poles at $s=4m^2$}} \nn \, ,
\ea
where the $\tau_i$ dependent coefficients $c^{s,t,u}_a$, $d^{s,t,u}_a$, $C_{ab}$ and $D_{ab}$ are dimensionless and naturally of order unity or vanishing. Additional constraints can be found by imposing crossing symmetry \cite{deRham:2017zjm,Bonifacio:2018vzv}, but will not concern us here. The above scalings are indicative of the fact that the helicity-2 poles come in at the scale $1/\mpl^2=(m^8 g_*^2)/(\Lambda_5^{10})$, however the helicity-1 and helicity-0 are additionally enhanced by up to four powers of $1/m^2$ from the unitary gauge polarization structure and external state normalization. Regardless, the leading interactions then enter at the scale $g_*^2/(\Lambda_5^{10})$ and higher dimension operators in the effective theory expansion will be further suppressed by positive integer powers of $1/\Lambda_5^2$, indicative of new states at this scale. \\

When applying the positivity bounds under the assumption that the UV completion is weakly coupled, we may take the threshold scale ${\cal M}^2 \sim \Lambda_5^2$. Given the above form for the amplitude, the higher derivative EFT corrections which come in at ${\cal O}\(g_*^2/\Lambda_5^{12} \)$ are $1/\Lambda_5^2$ suppressed relative to the leading interactions, and these are of the same order as the $1/{\cal M}^2$ terms in the positivity bounds \eqref{eq:2ndbound} that arise from the leading amplitude. \\

Focusing on the leading order ${\cal O}\(g_*^2/\Lambda_5^{10} \)$ interactions, then given a large hierarchy $\Lambda_5 \gg m$, the positivity bounds may be simplified to
\ba
\label{possimpl}
&& \frac{\partial^{2N}}{\partial v^N} f_{\alpha \beta} \Big|_{t=v=0}>0  \, ,  \quad \forall  N \ge 0  \, ,  \, \, {\text{  and}}\\
\label{possimpl2}
&& \frac{\partial^{2N}}{\partial v^N} \frac{\partial^M}{\partial t^M} f_{\tau_1 \tau_2} \Big|_{t=v=0}>0 \,  , \quad \forall M \ge 1, N \ge 0\,,
\ea
for all those $N$ and $M$ for which the LHS is non-zero, together with
\ba
\label{possimpl3}
&&f_{\ti_1\ti_2}(v,t) >0, \, \quad \text{ for } |v| \ll  \Lambda_5
\\
\label{possimpl4}
&&\frac{\pd}{\pd t} f_{\ti_1\ti_2}(v,t)   >0 , \, \quad \text{ for } |v| \ll  \Lambda_5\,,
\ea
plus higher $t$ derivatives. In practice from the leading amplitude, there are only a finite number of $v$ derivatives which are non-zero. For the $t$ derivatives there are an infinite number nonzero provided $N=0$ due to the $t$-channel pole. However in practice, beyond the first few $t$ derivatives, higher order ones will be dominated by the $t$ channel pole, and will eventually degenerate into the statement that the residue of the $t$-channel pole must be positive.  \\

We thus conclude, that it is consistent to first truncate the scattering amplitude to the leading ${\cal O}\(g_*^2/\Lambda_5^{10} \)$ interactions and to impose the simplified positivity bounds \eqref{possimpl}, \eqref{possimpl2}, \eqref{possimpl3}, \eqref{possimpl4} to the scattering amplitude for those finite number of terms that give non-trivial independent information. Once this has been done, the $ {\cal O}\(g_*^2/\Lambda_5^{12} \)$ EFT corrections may be included which compete with the $1/{\cal M}^2$ suppressed terms from the leading interactions in \eqref{eq:2ndbound}. The bounds should then be applied for only those terms for which the leading interactions contributed zero. Repeating this process, nontrivial bounds may be applied to the coefficients of the interactions to any order in the EFT expansion.

%%%%
\paragraph{Scattering Amplitudes:}
%%%%

Following the above discussion, we may first consider only the leading interactions that come from the mass potential, which may in unitary gauge be written in the form:
\begin{align}
\mathcal{L} &\supset  \frac{\mpl ^2}{2}  \left( R \left[ g \right] - \frac{m^2}{4} V (g, h) \right)\,.
\end{align}
In order to compare with previous works we further parameterize the interactions to quartic order in the manner\footnote{
There are a few parametrization for the mass terms. The relation between $c_3$ and $d_5$ and $\alpha_3$ and $\alpha_4$ is given right after Eq.~(23) of \cite{deRham:2010kj}: $\alpha_3=-2c_3$ and $\alpha_4=-4 d_5$. The relation between $\ai_i$ and $\bi_i$ can be found, for instance, in Eq.(6.24) of \cite{deRham:2014zqa}.
}
 \begin{align}
\label{eqn:GravityLeading}
V(g,h) \supset & [ h^2 ] - [ h ]^2 + (c_1 - 2 ) [ h^3 ] + (c_2 + \frac{5}{2} ) [ h^2] [h]  \\
&+ (d_1 +3 - 3 c_1) [ h^4] + (d_3 - \frac{5}{4} - c_2 ) [ h^2]^2 +... ~.\nn
\end{align}
Here $[h] = \eta^{\mu\nu} h_{\mu\nu}$, $[h^2] = \eta^{\mu\nu} h_{\mu \alpha} \eta^{\alpha \beta} h_{\beta \nu}$, etc.. The expected order of magnitude for the coefficients $c_1, c_2, d_1 , d_3$ can be determined by matching in unitary gauge to the expansion of the action \eqref{eq:Lambda5_1} or \eqref{eq:Lambda5_2}. In order to bridge comparison with previous treatments and the $\Lambda_5$ theory we shall however continue to remain agnostic about their magnitude. The fluctuations are then canonically normalized by performing the redefinition $h_{\mu \nu} \to 2 h_{\mu \nu}/\mpl $ so that the propagator is
\begin{align}
D_{\mu \nu \alpha \beta} (p) = \frac{ 1 }{p^2 +m^2} \left(  \frac{1}{2} \Pi_{\mu \alpha} \Pi_{ \nu \beta } + \frac{1}{2} \Pi_{\mu \beta} \Pi_{ \nu \alpha }  - \frac{1}{3} \Pi_{\mu \nu} \Pi_{\alpha \beta}    \right)  , \;\;\;\; \Pi_{\mu \nu} = \eta_{\mu \nu} + \frac{ p_\mu p_\nu }{m^2} .
\end{align}
It is convenient to define
\begin{equation}
 d_3 = -d_1/2 + 3/32 + \Delta d , \;\;\;\; c_2 = -3c_1/2 + 1/4 + \Delta c    ,
\label{eqn:L5Coeff}
\end{equation}
and interpret the bounds on the parameter space $\{ c_1, d_1, \Delta c, \Delta d  \}$. $\Delta c = \Delta d = 0$ corresponds to the $\Lambda_3$ massive gravity tuning which results in the $\Lambda_3$ theory to be considered later.

%%%%%%%%%%%%%%%%
\paragraph{Polarizations:}
%%%%%%%%%%%%%%%%
To exploit crossing symmetry, it is helpful to work in the transversity basis. For momenta $k^\mu = (\omega, 0,0, k)$, the corresponding polarizations are,
\begin{align}
\epsilon_{\mu \nu}^{(\tau = \pm 2)} &= \frac{1}{2m^2} \left(
\begin{array}{cccc}
 k^2 & \pm i k m & 0 & k w \\
 \pm i k m & -m^2 & 0 & \pm i m w \\
 0 & 0 & 0 & 0 \\
 k w & \pm i m w & 0 & w^2 \\
\end{array}
 \right) ,
\qquad
\epsilon_{\mu \nu}^{(\tau = \pm 1)} = \frac{1}{2m} \left(
\begin{array}{cccc}
 0 & 0 & i k & 0 \\
 0 & 0 & \mp m & 0 \\
 i k & \mp m & 0 & i w \\
 0 & 0 & i w & 0 \\
\end{array}
\right) ,
\nonumber \\
\epsilon_{\mu \nu}^{(\tau=0)} &= \frac{1}{\sqrt{6} m^2}
\left(
\begin{array}{cccc}
 k^2 & 0 & 0 & k w \\
 0 & m^2 & 0 & 0 \\
 0 & 0 & -2 m^2 & 0 \\
 k w & 0 & 0 & w^2 \\
\end{array}
\right)\,,
\end{align}
and we can express a general spin state via a five component vector $\alpha$,
\begin{equation}
\epsilon_{\mu \nu}^{(\alpha)} = \sum_{\tau} \alpha_\tau \epsilon_{\mu \nu}^{(\tau)} .
\end{equation}
These polarizations are related to the standard $SVT$ decomposition by
\begin{equation}
\arraycolsep=6.0pt\def\arraystretch{1.2}
\left( \begin{array}{c}
\alpha_{T_1} \\
\alpha_{T_2} \\
\alpha_{V_1} \\
\alpha_{V_2} \\
\alpha_{S}
\end{array} \right)  = \frac{1}{2 \sqrt{2}} \left( \begin{array}{c c c c c}
- 1 & 0 & \sqrt{6} & 0 & - 1 \\
0 & 2 & 0 &  - 2  & 0 \\
- 2 & 0 & 0 & 0 & 2 \\
0 & 2 & 0 & 2 & 0 \\
\sqrt{3}  & 0 & \sqrt{2} & 0 & \sqrt{3}
\end{array}\right) \left( \begin{array}{c}
\alpha_{-2} \\
\alpha_{-1} \\
\alpha_{0} \\
\alpha_{+1} \\
\alpha_{+2}
\end{array} \right)\,.
\end{equation}
It is more useful to express the residues $f_{\ai\bi}$ in terms of $\alpha_{S,V,T}$ because these polarizations have definite scaling with $s$
\be
\label{eq:highenergy}
\epsilon^{(T)} \sim s^0 , \;\;\; \epsilon^{(V)} \sim \frac{s}{m} , \;\;\; \epsilon^{(S)} \sim \frac{s^2}{m^2} ,
\ee
and correspond more closely to scattering $\phi$, $A$ or $h$ \stu fields.

%%%%%%%%%%%%%%%%
\paragraph{Forward Limit:}
%%%%%%%%%%%%%%%%
We define the positive residue
\be f_{\tau_1 \tau_2} = \frac{1}{10!} \frac{\partial^{10}}{ \partial s^{10}} \left[  s^4 (s-4m^2)^{4} \left( T_{\tau_1 \tau_2 \tau_1 \tau_2} (s, \theta) + T_{\tau_1 \tau_2 \tau_1 \tau_2}  (s, -\theta)      \right)     \right]  \, ,
\ee
as described in Section \ref{sec:positivity}. We will explore bounds provided by $f_{\tau_1 \tau_2}$ in what follows, but first we consider the bound inferred by imposing indefinite transversity $\frac{\partial^2}{\pd v^2} f_{\alpha \beta} >0$ in the forward limit as it is allows us to restrict the parameter space. \\

In the forward limit, we have the leading order bound
\begin{align}
2 \mpl^2 m^6 \frac{\partial^2}{\pd v^2} f_{\alpha \beta} |_{t=0} =&
\frac{352}{9} | \alpha_S \beta_S |^2
\left( \Delta c \left( -6 + 9 c_1 - 4 \Delta c  \right) - 6 \Delta d  \right)  \nn  \\
&+
\frac{176}{3}  \alpha_S^* \beta_S^* ( \alpha_{V_1} \beta_{V_1} -  \alpha_{V_2} \beta_{V_2}  )
 \,  \Delta c \left( 3 - 3 c_1 + 4 \Delta c  \right)  .
\end{align}
where $\alpha_\tau$ and $\beta_\tau$ are purely real\footnote{
Considering complex $\alpha_\tau$ and $\beta_\tau$ does not yield stronger bounds, so for brevity we shall quote the real expressions.
}.  Significantly, there exists a choice of polarizations, namely,
\begin{equation}
\alpha_S = \epsilon , \;\; | \alpha_{T_1} |^2 + |\alpha_{T_2}| ^2 = 1 - \epsilon^2 - | \alpha_{V_1} |^2 - | \alpha_{V_2} |^2\,,
\end{equation}
such that
\begin{equation}
2\mpl^2 m^6 \frac{\partial^2}{\pd v^2} f_{\alpha \alpha} |_{t=0} = \frac{176}{3} \left( \alpha_{V_1} \alpha_{V_1} -  \alpha_{V_2} \alpha_{V_2} \right)  \Delta c \left( 3 - 3 c_1 + 4 \Delta c  \right) \left( \epsilon^2 + \mathcal{O} ( \epsilon^4 ) \right)\,.
\end{equation}
This must be positive for all values of $\alpha_{V_1}$ and $\alpha_{V_2}$ (with $| \alpha_{V_1} |^2 + | \alpha_{V_2} |^2 \leq 1$), and therefore one is forced to set
\begin{equation}
\Delta c = 0
\end{equation}
to this order\footnote{
In principle, it could be $\mathcal{O} (1/\mpl )$ operators in such a way that higher derivative operators are capable of satisfying the bound.
}, which further imposes $\Delta d \le 0$. Remarkably one of the $\Lambda_3$ massive gravity tunings which raises the cutoff from $\Lambda_5$ to $\Lambda_3$ is then forced on us by the positivity bounds. \\

The other forward limit bound is quite cumbersome to display, but can be written more succinctly by noting that only certain combinations of the polarization $\alpha_{\tau}$ may appear (while respecting particle exchange and parity invariance). Specifically using the definitions in Appendix \ref{sec:ampdefs} , then we have
\begin{align}
2 m^2 \mpl^2 f_{\alpha \beta} |_{t=0} =&
2 \alpha_T^2 \beta_T^2	
+ X_S^2
			\left( \frac{55}{18} + \frac{10}{3} c_1 - 2 c_1^2 - \frac{32}{9} d_1  + \frac{32}{3} \Delta d ( 2 - 11 \frac{v^2}{m^4} )      \right)
			 \\
&+ X_{V_+}^2
			\left(  - \frac{7}{2} + 12 c_1  - \frac{15}{2} c_1^2 - 16 \Delta d  \right)
+ X_{V_-}^2
			\left( 6 - 6 c_1 + \frac{9}{2} c_1^2 - 4 d_1  \right)
			\nn \\
& + X_{SV}
			\left( 8 - 9 c_1 + \frac{9}{2} c_1^2  - \frac{8}{3} d_1   \right)
+ X_S X_{V_+}			
			\left(  18 - 38 c_1 + 21 c_1^2  \right)
			\nn \\
&+ X_{ST}
			\left(  \frac{16}{3} - 4 c_1   \right)
+ X_S X_T
			\left( - \frac{52}{3} + 32 c_1 -24 c_1^2  + \frac{32}{3} d_1  - \frac{64}{3} \Delta d  \right)
		\nn \\
&+ X_{V_+} X_T
			\left( 12 - 24 c_1 + 12 c_1^2   \right)
+ X_{VT}			
			\left( 4 - 3 c_1  \right)
			\nn \\
&+ \sqrt{3} X_{STVV}
			\left(  \frac{4}{3} - 2 c_1  + 3 c_1^2 - \frac{8}{3} d_1   \right)
- \frac{1}{\sqrt{3}} X_{SVVT}			
			\left( 3 c_1^2 - 2 \right)^2  \nn  .
\end{align}

\begin{figure}
\includegraphics[height=0.37\textwidth]{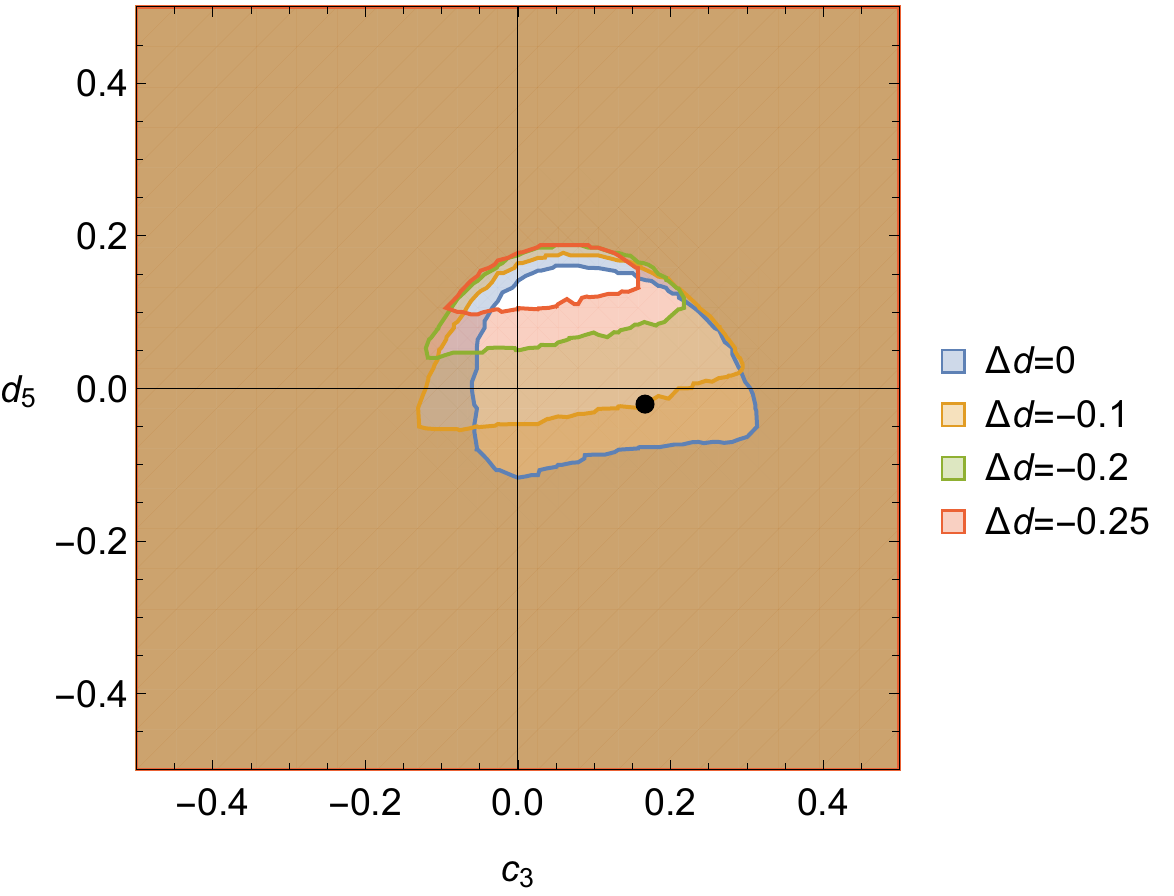}
\includegraphics[height=0.37\textwidth]{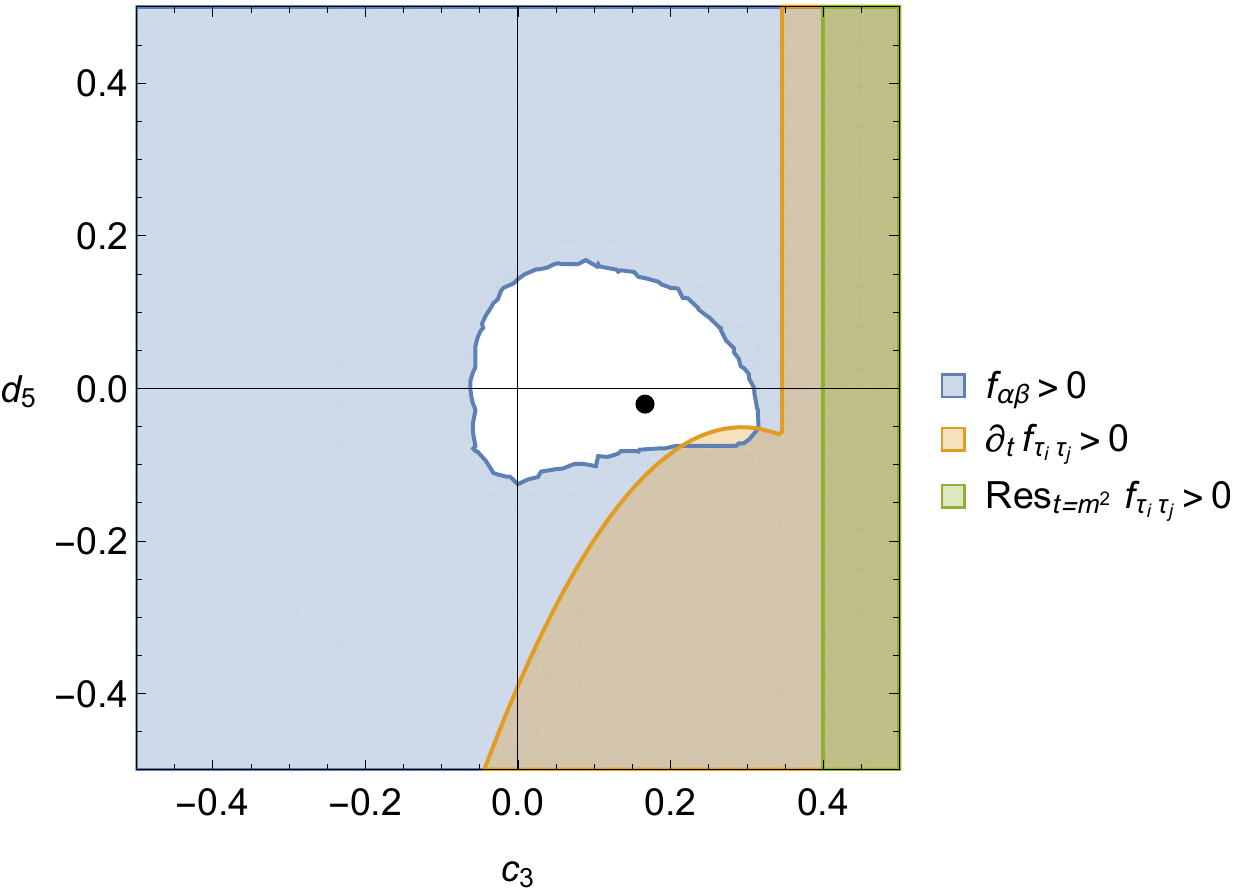}
\caption{Parameter space of massive gravity constrained by analyticity. The forward limit bounds on $\Lambda_5$ massive gravity (left) gradually constricts as $\Delta d$ is made more negative. However, going beyond the forward limit \eqref{eqn:L5dtbound} rules out $\Delta d \neq 0$ completely. In $\Lambda_3$ massive gravity (right), the higher $t$ derivative bounds marginally restrict the forward limit bounds. This lends further evidence to the idea that $\Lambda_3$ massive gravity may admit a Wilsonian UV completion within this narrow island.
\label{fig:L5}}
\end{figure}

Note that a negative $\Delta d$ can relax the bounds imposed by $SS$, $V_1 V_1$ and $V_2 V_2$ scattering. Finding the analytic minimum of this expression (a quartic form in $\alpha_\tau \beta_\tau \alpha_\tau^* \beta_\tau^*$) is an NP hard problem \cite{Cheung:2016yqr}, so we present an allowed region of parameter space which is found by approximate numerical minimization (see Figure~\ref{fig:L5}). \\

We see that the forward limit positivity requirements on the four point function require that the coefficients $c_1, c_2$ are tuned to the special $\Lambda_3$ massive gravity values, but $d_1$ and $d_3$ may so far differ. In particular, minimizing the bound numerically, it is found that when considering the leading order bound alone in the forward limit, then analyticity prefers a large negative $\Delta d$. This situation is changed dramatically when we consider the $t$ derivative bounds as we shall see below.

%%%%%%%%%%%%%%%%
\paragraph{First $t$ derivatives:}
%%%%%%%%%%%%%%%%

The leading $s^5$ contribution gives
\begin{equation}
2 \mpl^2 m^8 \partial_t \partial_v^2 f_{\alpha \beta}\quad  \propto \quad \Delta c^2 \;| \alpha_S |^2 | \beta_S |^2    ,
\end{equation}
which vanishes to this order when we take $\Delta c = 0$ to satisfy the forward limit bounds. (Note that the $t$ derivative bounds only apply for definite transversity; here the use of $f_{\alpha \beta}$ is only for book-keeping, i.e., to write the various independent $f_{\ti_1 \ti_2}$ quantities in more compact way.) The other $t$ derivative bound may be written as
\begin{align}
2 \mpl^2 m^4  \frac{\partial}{\pd t} f_{\ai\bi} \Big|_{t=0} =&
+ 2 \alpha_T^2 \beta_T^2			
+ X_{V_+}^2
			\left(  \frac{41}{4}  - \frac{33}{2} c_1  + \frac{27}{4} c_1^2   \right)
+ X_{V_-}^2			
			\left( 8 -12 c_1 + \frac{9}{2} c_1^2   \right)
			 \\
&+ X_S^2
			\left(  \frac{925}{36} - 43 c_1 + 21 c_1^2  - \frac{32}{9} d_1 + \frac{32}{9} \Delta d ( -6 + 22 \frac{v}{m^2} )   \right)
+ X_{VT}			
			\left( 4 - 3 c_1  \right)
			\nn \\
&+ X_{ST}
			\left(  7 - 6 c_1  \right)
+ \sqrt{3} \alpha_S \beta_S \left( \alpha_S \beta_{T_1} + \beta_S \alpha_{T_1} \right)
			\left(  - \frac{4}{9}  + 2 c_1   - \frac{16}{9} d_1    +   \frac{32}{9} \Delta d     \right)
			\nn \\
&+ ( \alpha_S^2 \beta_{V_1}^2 + \beta_S^2 \alpha_{V_1}^2 )
			\left( \frac{40}{3}     - 21 c_1  + \frac{33}{4} c_1^2  - \frac{32}{3} \Delta d   \right)
			\nn \\
&+ ( \alpha_S^2 \beta_{V_2}^2 + \beta_S^2 \alpha_{V_2}^2 )
			\left(  \frac{44}{3}   - 23 c_1   + \frac{45}{4} c_1^2 - \frac{8}{3} d_1   \right)
			\nn \\
&+ \alpha_S \beta_S \alpha_{V_1} \beta_{V_1}
			\left(   \frac{101}{6}   - 33 c_1 + \frac{33}{2} c_1^2   - \frac{176}{3} \Delta d  \right)
			\nn \\
&+ \alpha_S \beta_S \alpha_{V_2} \beta_{V_2}
			\left(  \frac{43}{6}    - 11 c_1 + \frac{27}{2} c_1^2  - \frac{32}{3} d_1  + 16 \Delta d    \right) \nn \,.
\end{align}
These tree-level amplitudes can be used in the positivity bounds with $\mathcal{M}^5 \sim \mpl m^4= \Lambda_5^5$ as the cutoff, and as we have discussed for the leading interactions it is consistent to take the bounds \eqref{possimpl}-\eqref{possimpl3}
\ba
 && \frac{\partial}{\pd t} f_{\ti_1\ti_2}  (v, t) > 0  ,  \\
 && \frac{\partial^3}{\pd t\pd v^2}  \left[ f_{\ti_1\ti_2} (v,t)  \right] > 0 .
\ea
The latter simply sets $\Delta c =0 $ as before. Assuming a hierarchy between $m^2$ and $\mu_b \gtrsim \Lambda_5$ (the scale at which the branch cut begins), we can consider $|v|$ in the range $m^2 \ll |v| \ll \mu_b$, and the first $t$ derivative bound gives,
\be
\frac{\partial}{\pd t} f_{\ti_1\ti_2}  (v, t) \;\;  \propto \;\;  \frac{m^2 v}{ \Lambda_5^{10}}  \Delta d + \mathcal{O} \left( \frac{m^4}{\Lambda_5^{{10}}}  \right)    \;\;  > \;\;  0\,.
\label{eqn:L5dtbound}
\ee
As $v$ can take either sign, this enforces the condition,
\be
\Delta d = 0 .
\ee
Of the parameters which appear in the 2-to-2 scattering amplitude, analyticity requires the special $\Lambda_3$ tuning.

%%%%%%%%
\paragraph{Goldstone Equivalence:}
%%%%%%%%

The bound which forces $\Delta d = 0$ beyond the forward limit is from $SS$ scattering. While scattering indefinite transversities do not obey positivity conditions beyond the forward limit, this particular combination,  does because it has trivial crossing properties at high energies
\be
 \epsilon_S =  \frac{1}{2} \epsilon_{\tau=0} + \frac{\sqrt{6}}{4} \left(  \epsilon_{\tau=+2} + \epsilon_{\tau=-2}  \right) \, .
\ee

Significantly, while the forward limit bound strengthens gradually as $\Delta d$ is made more negative (imposing $-0.3 \lesssim \Delta d \leq 0$), as shown in Fig.~\ref{fig:L5}, the first $t$ derivative bound imposes the much stricter requirement that $\Delta d = 0$.
It is this analyticity result that makes raising the cutoff from $\Lambda_5$ a well-motivated thing to do in the massive spin-2 EFT, supposing that the theory had come from an underlying, analytical, local, Lorentz-invariant UV completion, then the na\"{i}vely eight-dimensional parameter space in $ \{ c_i, d_i \} $ is (at least partially) projected onto $\Lambda_3$ massive gravity.

%%%%%%%%%%%%%%%%%%%%%%%%%%%%
\section{$\Lambda_3$ Massive Gravity}
\label{sec:L3}
%%%%%%%%%%%%%%%%%%%%%%%%%%%%

In this section, we consider the consequences of the 'ghost-free massive gravity' tuning \cite{deRham:2010ik,deRham:2010kj} which raises the cutoff of massive gravity from $\Lambda_5$ to $\Lambda_3$.

%%%%%%%%%%%%%%%%
\paragraph{Raising the cutoff:}
%%%%%%%%%%%%%%%%

As we have discussed, generic massive gravity has a unitarity cutoff at $\Lambda_5 = (m^4 \mpl )^{1/5}$. In the EFT construction, this manifests itself as an $SSSS$ which scales as $s^5/\Lambda_5^{10}$. This cutoff can be raised as high as $\Lambda_3 = (m^2 \mpl )^{1/3}$ by performing the tuning
\begin{equation}
c_1 = 2 c_3 + \frac{1}{2}, \;\; c_2 = - 3 c_3 - \frac{1}{2} , \;\;\; d_1 = -6 d_5 + \frac{3}{2} c_2 + \frac{5}{16} , \;\;\; d_3 = 3 d_5 - \frac{3}{4} c_3 - \frac{1}{16}   .
\label{eqn:dRGTcond}
\end{equation}
This removes the $s^5$ and $s^4$ contributions to every four-point function at tree-level.
As discussed earlier,  performing these tunings is \emph{not}, a priori, a particularly natural thing to do in a bottom-up approach of EFT, although it is technically natural (see  discussions in \cite{deRham:2012ew,deRham:2013qqa}). However, we have seen that analyticity/positivity requires that the coefficients are tuned precisely in this manner. This is a novel and unexpected result, which could not have been derived from considerations of the LEEFT alone.

%%%%%%%%%%%%%%%%
\paragraph{Forward Limit:}
%%%%%%%%%%%%%%%%

The forward limit bounds can be written in the form
\begin{equation}
2 \mpl^2 m^2 f_{\ai\bi} |_{t=0} = n_0 + n_1 c_1 + n_2 c_2^2 + n_3 d_5  > 0    ,
\end{equation}
where the strictest $n_i$ coefficients are found by scanning over all possible $\alpha_\tau, \beta_\tau$ polarizations.
The most restrictive directions are approximately:
\begin{align}
\begin{array}{c c c c c | c c c c}
\alpha_S, \beta_S  &  \alpha_{V_1}, \beta_{V_1} & \alpha_{V_2}, \beta_{V_2} & \alpha_{T_1}, \beta_{T_1} & \alpha_{T_2}, \beta_{T_2} & n_0 & n_1 & n_2 & n_3  \\
\hline
\frac{\sqrt{3}}{\sqrt{8}}, \frac{\sqrt{3}}{\sqrt{8}} & \frac{1}{\sqrt{2}}, - \frac{1}{\sqrt{2}} & 0,0 & -\frac{1}{\sqrt{8}},-\frac{1}{\sqrt{8}} & 0,0 & \frac{11}{8} & \frac{51}{8} & -\frac{273}{8} & 0  \\
0,0 & 1,1 & 0,0 & 0,0 & 0,0 & \frac{5}{8} &  9 & -30 & 0  \\
0,0 & 1,0 & 0,1 & 0,0 & 0,0 & \frac{23}{8} &  -9 & 18 & 24  \\
\frac{1}{2},  \frac{1}{2} & 0,0 & \frac{1}{2},  \frac{1}{2} & \frac{2}{\sqrt{8}}, \frac{2}{\sqrt{8}} & 0,0 & 1.423 &  5.621 & -30.72 & -9.566  \\
\frac{\sqrt{2750}}{100} ,  \frac{\sqrt{2750}}{100} & \frac{\sqrt{2350}}{100}, - \frac{\sqrt{2350}}{100} & \frac{\sqrt{4500}}{100}, \frac{\sqrt{4500}}{100} & -\frac{\sqrt{50}}{100}, -\frac{\sqrt{50}}{100} & \frac{\sqrt{350}}{100}, -\frac{\sqrt{350}}{100} & 1.327 & 5.039 & -20.13 & 10.92
\end{array}
\label{eqn:L3Leading}
\end{align}
The first two rows restrict $c_3$ to the interval
\begin{equation}
-0.0582 \approx \frac{9-2\sqrt{39}}{60} < c_3 < \frac{51 + \sqrt{14613} }{546} \approx 0.315   ,
\end{equation}
while the next two rows restrict $d_5$ to a finite region
\begin{equation}
\frac{1}{24} \left( - \frac{23}{8} + 9 c_3 - 18 c_3^2  \right)  < d_5 \lesssim 0.149 + 0.588 c_3 + 3.21 c_3^2   .
\end{equation}
The final row demonstrates the small improvements which can be achieved by considering increasingly complicated superpositions, which rule out the lower left corner of the island:
\begin{equation}
\text{If }  -0.0582 \lesssim c_3 \lesssim 0   ,\;\;\;\; \text{ then } \;\;\;\; d_5 > -0.122 - 0.461 c_3 + 1.84 c_3^2  .
\end{equation}\\

%%%%%%%%
\paragraph{First $t$ derivative:}
%%%%%%%%

The bound from $\partial f_{SS}/\pd t$ has already been plotted in Figure~\ref{fig:L5} ($\Delta d = 0$ corresponds to $\Lambda_3$ massive gravity). Significantly, we see that even with the $\Lambda_3$ massive gravity tuning, one still gains a new constraint on the parameter space by studying the first $t$ derivative -- it rules out the lower right hand corner of the forward limit island. Explicitly, this is the $\partial_t f_{SS}$ bound,
\be
25 + 4 c_3 (-37 + 63 c_3) + 64 d_5 > 0\,.
\ee

%%%%%%%%
\paragraph{Higher $t$ derivatives:}
%%%%%%%%

Taking higher $t$ derivatives picks out the residue of the $t$ channel pole, which is,
\be
 \text{Res}_{t=m^2} \left[ f_{\alpha \beta} \right]  = \frac{1}{4}  Y_\alpha Y_\beta  +  12 (c_1 - 1 )^2 \alpha_S \alpha_{V_1} \beta_S \beta_{V_1}   \, ,
\ee
for the combinations,
\begin{equation}
Y_\alpha = -2 \alpha_T^2 + (3c_1 - 4 ) \alpha_V^2 + (6c_1 - 7) \alpha_S^2  .
\end{equation}
This must be positive for any definite transversity scattering. The strongest such bound is from scattering $\tau=0$ with $\tau = \pm 2$, which yields,
\begin{equation}
\( c_3 - \frac{2}{5} \) \(c_3 - \frac{5}{6} \)  > 0 .
\end{equation}
As the other positivity bounds have already restricted us to $c_3$ below $\frac{2}{5}$, this does not yield any new information.
As shown in Figure~\ref{fig:L5}, if the EFT satisfies the leading forward and first $t$ derivative bounds, then it will satisfy all higher $t$ derivative bounds.

%%%%%%%%%%%%%%%%%%%%%%%%%%%%
\section{Discussion}
\label{sec:conc}
%%%%%%%%%%%%%%%%%%%%%%%%%%%%

Recent advances in positivity bounds---requirements which low energy  coefficients must satisfy in order for compatibility with a Wilsonian UV completion---provide powerful new tools with which to probe and constrain the parameter space of gapped EFTs. In this article we have discussed the important case of bounds applied to the low energy EFTs of massive spin-1 and spin-2 particles with the assumption that the UV completion is weakly coupled. This allows us to place the bounds on the tree-level scattering amplitudes and in turn on the coefficients in the tree-level Wilsonian effective action.\\

We have demonstrated how the introduction of \stu fields facilitates a straightforward power counting for the bottom-up construction of EFTs for massive spinning particles, in the particular cases of spin-1 (Proca) and spin-2 (massive gravity) (see also \cite{deRham:2017xox}). Certain tunings of the EFT coefficients are stable under quantum corrections, examples of which are the charged Galileon (obtained by tuning the leading order coefficient in the Proca EFT), and $\Lambda_3$ massive gravity (obtained by raising the cutoff of $\Lambda_5$ massive gravity). \\

The positivity constraints have been derived for these theories, and it is found that the $t$ derivative bounds provide genuinely orthogonal information to the forward limit bounds alone.
For example, it is found that massive gravity can admit no local, Lorentz invariant UV completion unless the parameters are tuned, at least to quartic order, to the 'ghost-free' massive gravity structure which raises the strong coupling scale above to $\Lambda_3$. This is a novel result, and demonstrates the power of these bounds in extending beyond naive EFT expectations. The leading order bounds presented in this work are summarised in Table~\ref{tab:summary}. \\

We stress that many more bounds are available as one goes to higher and higher orders in the EFT/energy expansion, and that in principle it is possible to constrain the new coefficients which appear at every order (not just leading order). The leading order calculations presented here are not only valuable for potential cosmological applications of these models, they are also an important proof of principle: that positivity bounds provide tight constraints on the parameter space of spinning particles, and that going beyond the forward limit yields new information, independent of the forward limit, and independent of the Goldstone/decoupling limit in which the bounds are applied only to the Goldstone modes.\\

\begin{table}
\centering
\begin{tabular}{c | c c}
				& Forward bound 	& First $t$ derivative  \\
				\hline\vspace{3pt}
Proca		&   $a_0 > 0 $								& $6 \bar{a}_t + 112 a_0 > 0 $ \\
				& 	$c_1 > \text{Min} [ 0, -2c_2 ]$ 	&   \\[3pt]
				\hline\vspace{3pt}
Charged Galileon &  $ a_0 + \frac{1}{2} a_3 - \frac{1}{2} a_4 > 0  , \;\; | \mu_3 | < X $   &  $\bar{a}_t > 0$ \\
				&   $c_1 > \text{Min} [ 0 , - 2 c_2 ]$  &   \\[3pt]
				\hline\vspace{3pt}
$\Lambda_5$ Massive Gravity  & $ \Delta c = 0, \; \Delta d \lesssim 0 $  & $\Delta d  =0$  \\[3pt]
				\hline \vspace{3pt}
$\Lambda_3$ Massive Gravity  & $-0.058 c_3 \lesssim 0.32 $  & (see Fig.~\ref{fig:L5}) \\
&  $n_3 d_5 > - n_0 - n_1 c_3 - n_2 c_3^2 $  &   \\
\end{tabular}
\caption{Summary of the leading order positivity bounds for massive spin-1 and spin-2 EFTs. See equations (\ref{eqn:ProcaLeading}, \ref{eqn:Proca_mu}, \ref{eqn:GravityLeading}, \ref{eqn:L5Coeff}, \ref{eqn:L3Leading}) for definitions of these EFT coefficients. The $\Lambda_3$ bounds can be improved numerically by scanning indefinite combinations of transversities in the forward limit. The forward limit bounds on the charged Galileon compare favourably with those of \cite{Bonifacio:2016wcb}, as do the forward limit bounds on $\Lambda_3$ massive gravity with \cite{Cheung:2016yqr}.
\label{tab:summary}}
\end{table}

\paragraph{UV completion and Higgs mechanism:}The spin-1 theories we have discussed can be understood as the low energy effective theories in a heavy Higgs mechanism in which the heavy Higgs particle is integrated out. Similarly the massive spin-2 theories can be viewed as the low energy effective theories of some equivalent mechanism which spontaneously breaks the diffeomorphism symmetry. The absence of a Higgs mechanism for spin-2 has been recently argued in \cite{Arkani-Hamed:2017jhn}. More precisely these authors note that in the high energy limit $E \gg m$, the coupling of the helicity zero and one modes of the massive graviton with a single helicity two are not of the same strength as implied by the equivalence principle \cite{Weinberg:1964ew}. This is a well known result, the interactions of the helicity scalar and vector modes do not package together into diffeomorphism scalar and vectors. Indeed the helicity vector arises from a diffeomorphism scalar in the \stu form of massive gravity. Perturbatively this is reflected by the fact that under a linear spin-2 gauge transformation $h_{\mu\nu} \rightarrow h_{\mu\nu} + \partial_{\mu} \xi_{\nu} + \partial_{\nu} \xi_{\mu}$, the vector transforms as $A_{\mu} \rightarrow A_{\mu} + m \xi_{\mu}$, in stark contrast to a massless spin-1 theory where the vector is invariant. The implication is that for $m \neq 0$, terms coupled linearly to $h_{\mu \nu}$ do no not have to be conserved, but only cancel against those terms coupling to $A_{\mu}$, i.e.
\be
S_{\rm int}=\int \d^4 x \frac{1}{\mpl} \( h_{\mu\nu} T^{\mu\nu}+ A_{\mu} J^{\mu} \)
\ee
transforms as
\be
\delta_{\xi} S_{\rm int}=\int \d^4 x \frac{1}{\mpl} \(  (\partial_{\mu} \xi_{\nu} + \partial_{\nu} \xi_{\mu})T^{\mu\nu}+ m \xi_{\mu} J^{\mu} \) =\int \d^4 x \frac{1}{\mpl}   \xi_{\mu} \(-2 \partial_{\nu} T^{\mu\nu} + m J^{\mu} \) \, .
\ee
At this level we may be tempted in ignoring the $J^\mu$ current as $m\to 0$ and infer $\p_\mu T^{\mu
\nu}=0$. However we {\it cannot} directly set $m=0$ at this level since the $\Lambda_3$ or $\Lambda_5$ interactions will diverge and there is a delicate interplay between positive powers of $1/m$ in the interactions and powers of $m$ in the gauge transformations and so in particular
\be
\lim_{m \rightarrow 0,\,  \mpl \text{  finite}} \(-2 \partial_{\nu} T^{\mu\nu} + m J^{\mu} \) =0  \text{ does not imply }\(-2 \partial_{\nu} T^{\mu\nu}  \)=0\,.
\ee
Weinberg's proof of the equivalence principle \cite{Weinberg:1964ew} no longer applies since the external scattering states are massive, and even in the high energy limit $E \gg m$ the naively small departures from masslessness are made significant by the powers of $1/m$ in the interactions. \\

This subtlety in understanding the high energy limit is just another realization of the vDVZ discontinuity \cite{vanDam:1970vg,Zakharov:1970cc} and is not present for $S \le 1$. Massive spin-2 particles can realize the equivalence principle, in the sense that there is a universal gauge transformation, as above, even though the graviton/scalar/vector interactions do not follow the form implied by \cite{Weinberg:1964ew}, since the latter only applies for true massless scattering states.  The correct way to understand the high energy limit $E \gg m$ is to take the decoupling limit $m \rightarrow 0$ with $\Lambda_3$ fixed. In this limit, the massless graviton couples to an identically conserved $T_{\mu\nu}$ consistent with massless spin-2 gauge invariance, and on-shell graviton-X-Y interactions arise at first order in the departure from the decoupling limit and are thus $m^2/\Lambda_3^2$ suppressed. But once again at this order, we must include the transformation of the vector $A_{\mu}$ under spin-2 gauge transformations and the on-shell graviton-X-Y interactions differ from \cite{Weinberg:1964ew}. \\

Thus while the argument of \cite{Arkani-Hamed:2017jhn} excludes the possibility that the states of a 'spin-1 like' Higgs mechanism in which a massless graviton combines with the states of a diffeomorphism scalar and vector to form a massive spin-2 particle, (which was never to be expected), it does not preclude an alternative as yet unknown symmetry breaking mechanism. In this sense, the positivity bounds we have discussed are a more reliable indicator of the possibility of a local, Lorentz invariant UV completion for massive spin-2 states and by extension an equivalent of the Higgs mechanism for spin-2 fields.

\vspace{12pt}
\noindent{\bf Acknowledgments}

We would like to thank Lavinia Heisenberg and David Pirtzkhalava for useful discussions and comments.
CdR and AJT would like to thank the Pauli Center for Theoretical Studies at the ETHZ\"urich for hosting them during the final part of work.
The work of CdR and AJT is supported by STFC grant ST/P000762/1. CdR thanks the Royal Society for support at ICL through a Wolfson Research Merit Award. CdR and SYZ are also supported in part by the European Union's Horizon 2020 Research Council grant 724659 MassiveCosmo ERC-2016-COG and CdR in part by a Simons Foundation award ID 555326 under the Simons Foundation's Origins of the Universe initiative, `{\it Cosmology Beyond Einstein's Theory}'. SM is funded by the Imperial College President's Fellowship. AJT thanks the Royal Society for support at ICL through a Wolfson Research Merit Award. SYZ acknowledges support from the starting grant from University of Science and Technology of China (KY2030000089) and the National 1000 Young Talents Program of China.\\

\appendix

%%%%%%%%%%%%%%%%%%%%%%%%%%%%
\section{Positivity Bounds for General Transversities}
%%%%%%%%%%%%%%%%%%%%%%%%%%%%

Following the notation and the proof of \cite{deRham:2017zjm}, for general transversities ($\ai_{\ti_1},\bi_{\ti_2}$), the $u$ crossing of
\be
\mathcal{T}^+_{\alpha \beta} = \sum_{\ti_1, \ti_2,\ti_3, \ti_4}  \ai_{\ti_1}\bi_{\ti_2} \ai_{\ti_3}^* \bi_{\ti_4}^* \mathcal{T}^+_{\ti_1\ti_2\ti_3\ti_4}(0,t)\,,
\ee
on the left-hand branch cut is given by
\begin{equation}
\text{Abs}_u \mathcal{T}^+_{\alpha \beta} = \sum_{\lambda_i, J, \nu}  z_{\lambda_1 \lambda_2}^{Ja} z^{Jb*}_{\lambda_3 \lambda_4} t_\nu^{(J \lambda \mu)} \cos \left( (a+b+\nu) \theta_u  \right)\,,
\end{equation}
where
\be
z^{Ja}_{\lambda_1 \lambda_2} =  \sum_{N, \tau_1, \tau_2} c^a_{\tau_1 \tau_2} (u) \alpha_{\tau_1} \beta_{\tau_2} u_{\lambda_1 \tau_1} u_{\lambda_2 \tau_2} \langle  N  | T | J M ,  \lambda_1 \lambda_2 \rangle\,.
\ee
If one can conclude that this left-hand branch cut is positive, one can then construct a standard dispersion relation for the scattering amplitude of four identical bosons of spin-$S$, $f_{\alpha \beta} (v, t)$, as defined in Eq.~(\ref{fvtdef}).
In the forward limit, $t=0$ ($\theta_u = 0$), this is indeed the case, because then we have
\be
 \sum_\nu t_\nu^{(J\lambda \mu)} = \delta_{\lambda \mu} \;\; \implies \;\; \text{Abs}_u \mathcal{T}^+_{\alpha \beta} \Big|_{t=0} = \sum_{\lambda_i J}  z_{\lambda_1 \lambda_2}^{Ja} z^{Jb*}_{\lambda_3 \lambda_4} \delta_{\lambda \mu}\,,
 \ee
and as $\delta_{\lambda \mu}$ is a positive definite matrix, the right-hand side is positive definite. This establishes that
\be
 \text{Abs}_u \mathcal{T}^+_{\alpha \beta}  \Big|_{t = 0} > 0\,,
\ee
and so if we set $t=0$ and expand in $v = s + t/2 - 2m^2$, we find a collection of positivity bounds:
\begin{align}
\partial_v^{2N} f_{\alpha\beta} (0,0)  > 0 \;\;\;\; \forall \;\;\;\; N  .
\end{align}

\section{Amplitude Definitions}
\label{sec:ampdefs}

For the purposes of writing down the spin-2 amplitudes it is convenient to define
\ba
&& \alpha_V^2 = \alpha_{V_1}^2 + \alpha_{V_2}^2  \,  ,   \\
&& \alpha_T^2 = \alpha_{T_1}^2 + \alpha_{T_2}^2  \,  , \\
&& X_S = \alpha_S \beta_S   \,  , \\
&& X_{V_+} = \alpha_{V_1} \beta_{V_1} - \alpha_{V_2} \beta_{V_2}  \,  , \\
&& X_{V_-} = \alpha_{V_1} \beta_{V_2} + \alpha_{V_2} \beta_{V_1}  \,  , \\
&& X_T = \alpha_{T_1} \beta_{T_1} - \alpha_{T_2} \beta_{T_2}   \,  , \\
&& X_{SV} = \alpha_S^2 \beta_{V}^2 + \beta_S^2 \alpha_V^2  \,  ,  \\
&& X_{ST} = \alpha_S^2 \beta_{T}^2 + \beta_S^2 \alpha_T^2  \,  , \\
&& X_{VT} =  \alpha_V^2 \beta_T^2 + \beta_V^2 \alpha_T^2   \,  ,  \\
&& X_{STVV} = \alpha_S \left(  \alpha_{T_1} ( \beta_{V_1}^2 - \beta_{V_2}^2 ) - 2 \alpha_{T_2} \beta_{V_1} \beta_{V_2} \right)  + \left(  \alpha \leftrightarrow \beta \right)  \,  ,  \\
&& X_{SVVT} = \alpha_S \left(  \alpha_{V_1} ( \beta_{V_1} \beta_{T_1} + \beta_{V_2} \beta_{T_2} ) +  \alpha_{V_2} ( - \beta_{V_1} \beta_{T_2} + \beta_{V_2} \beta_{T_1}  )  \right)  + ( \alpha \leftrightarrow \beta ) \, .
\ea

\section{Sufficient Conditions}
\label{App:sufficientCond}

The positivity bounds for the charged Galileon given in~\eqref{eq:fab2} is difficult to analytically minimize exactly. It can alternatively be written as,
\begin{align}
\frac{\Lambda_{\phi}^6}{8m^2} f_{\alpha \beta} \Big|_{t=0} &=
 \frac{4 \mu_2 + \mu_3}{8} \left(  | \alpha_+ \beta_0 - \alpha_0 \beta_+ |^2 +
 | \alpha_+ \beta_- + \alpha_- \beta_+ |^2 \right) \\
&+  \frac{ 4\mu_2 - \mu_3}{8} \left(  | \alpha_+ \beta_0 + \alpha_0 \beta_+ |^2 +
 | \alpha_+ \beta_-  -  \alpha_- \beta_+ |^2 \right) \nn \\
&+  \frac{ 3 \mu_4 - \mu_5 }{4} \left(  | \alpha_- \beta_- + \alpha_0 \beta_0 |^2 +
 | \alpha_0 \beta_- - \alpha_- \beta_0 |^2 \right) \nn \\
&+ \frac{\mu_4 + \mu_5}{4} | \alpha_0 \beta_- + \alpha_- \beta_0 |^2\nn  \\
&+ \frac{ \lambda_- }{  4  } |  \cos \frac{\theta}{2} \, \frac{\alpha_- \beta_- - \alpha_0 \beta_0}{\sqrt{2}}  - \sin \frac{\theta}{2} \,  \alpha_+ \beta_+ |^2  \nn \\
&+ \frac{\lambda_+ }{ 4 }  |   \sin \frac{\theta}{2} \, \frac{ \alpha_- \beta_- - \alpha_0 \beta_0}{\sqrt{2}}  +\cos \frac{\theta}{2} \, \alpha_+ \beta_+ |^2\nn\,,
\end{align}
where,
\begin{align}
\lambda_{\pm} &= 2 \mu_1 + 4 \mu_2 - \mu_4 + 3 \mu_5 \pm r \\
r \sin \theta &= \sqrt{2} \mu_3 , \;\; r \cos \theta = 2 \mu_1 + 4 \mu_2 + \mu_4 - 3 \mu_5\,.
\end{align}
If we imagine that all of the above mod squares can be varied independently, this gives a \emph{sufficient} condition for positivity,
\begin{equation}
| \mu_3 | < 4\mu_2, \;\;\;\; - \mu_4 < \mu_5  < 3 \mu_4 , \;\;\;\; \lambda_{\pm} > 0\,,
\end{equation}
which can be written as,
\begin{equation}
  \frac{1}{3} < \frac{\mu_5}{\mu_4}  < 3 , \;\;\;\; \mu_3^2 < \text{Min} \left[  16 \mu_2^2 , \,  4 ( \mu_1 + 2\mu_2) ( - \mu_4 + 3 \mu_5 )   \right] .
\end{equation}
This is overly restrictive because it is not possible to vary the $\lambda_{\pm}$ terms independently, i.e. a slightly negative $\lambda_-$ may be compensated by a sufficiently positive $\lambda_+$.

From these sufficient bounds, we see that $|\mu_3|<X$ where if is sufficient to have $X$  lies in the range
\ba
\text{Min} \left[  16 \mu_2^2 , \,  4 ( \mu_1 + 2\mu_2) ( - \mu_4 + 3 \mu_5 )   \right] < X < {\rm Min}\left[\mu_1+4\mu_2+\mu_5,\frac 12 (\mu_1+6\mu_2+4\mu_5)\right]\,.\qquad
\ea

%%%%%%%%%%%%%%%%%%%%%%%%%%%%
\bibliographystyle{JHEP}
\bibliography{refs}

\providecommand{\href}[2]{#2}\begingroup\raggedright\begin{thebibliography}{10}

\bibitem{Rattazzi:2008pe}
R.~Rattazzi, V.~S. Rychkov, E.~Tonni and A.~Vichi, \emph{{Bounding scalar
  operator dimensions in 4D CFT}},
  \href{http://dx.doi.org/10.1088/1126-6708/2008/12/031}{\emph{JHEP} {\bfseries
  12} (2008) 031}, [\href{https://arxiv.org/abs/0807.0004}{{\ttfamily
  0807.0004}}].

\bibitem{Caron-Huot:2016icg}
S.~Caron-Huot, Z.~Komargodski, A.~Sever and A.~Zhiboedov, \emph{{Strings from
  Massive Higher Spins: The Asymptotic Uniqueness of the Veneziano Amplitude}},
  \href{http://dx.doi.org/10.1007/JHEP10(2017)026}{\emph{JHEP} {\bfseries 10}
  (2017) 026}, [\href{https://arxiv.org/abs/1607.04253}{{\ttfamily
  1607.04253}}].

\bibitem{Paulos:2017fhb}
M.~F. Paulos, J.~Penedones, J.~Toledo, B.~C. van Rees and P.~Vieira, \emph{{The
  S-matrix Bootstrap III: Higher Dimensional Amplitudes}},
  \href{https://arxiv.org/abs/1708.06765}{{\ttfamily 1708.06765}}.

\bibitem{Adams:2006sv}
A.~Adams, N.~Arkani-Hamed, S.~Dubovsky, A.~Nicolis and R.~Rattazzi,
  \emph{{Causality, analyticity and an IR obstruction to UV completion}},
  \href{http://dx.doi.org/10.1088/1126-6708/2006/10/014}{\emph{JHEP} {\bfseries
  10} (2006) 014}, [\href{https://arxiv.org/abs/hep-th/0602178}{{\ttfamily
  hep-th/0602178}}].

\bibitem{Bellazzini:2016xrt}
B.~Bellazzini, \emph{{Softness and amplitudes’ positivity for spinning
  particles}}, \href{http://dx.doi.org/10.1007/JHEP02(2017)034}{\emph{JHEP}
  {\bfseries 02} (2017) 034},
  [\href{https://arxiv.org/abs/1605.06111}{{\ttfamily 1605.06111}}].

\bibitem{Cheung:2016yqr}
C.~Cheung and G.~N. Remmen, \emph{{Positive Signs in Massive Gravity}},
  \href{http://dx.doi.org/10.1007/JHEP04(2016)002}{\emph{JHEP} {\bfseries 04}
  (2016) 002}, [\href{https://arxiv.org/abs/1601.04068}{{\ttfamily
  1601.04068}}].

\bibitem{Bonifacio:2016wcb}
J.~Bonifacio, K.~Hinterbichler and R.~A. Rosen, \emph{{Positivity constraints
  for pseudolinear massive spin-2 and vector Galileons}},
  \href{http://dx.doi.org/10.1103/PhysRevD.94.104001}{\emph{Phys. Rev.}
  {\bfseries D94} (2016) 104001},
  [\href{https://arxiv.org/abs/1607.06084}{{\ttfamily 1607.06084}}].

\bibitem{deRham:2017avq}
C.~de~Rham, S.~Melville, A.~J. Tolley and S.-Y. Zhou, \emph{{Positivity bounds
  for scalar field theories}},
  \href{http://dx.doi.org/10.1103/PhysRevD.96.081702}{\emph{Phys. Rev.}
  {\bfseries D96} (2017) 081702},
  [\href{https://arxiv.org/abs/1702.06134}{{\ttfamily 1702.06134}}].

\bibitem{Pennington:1994kc}
M.~R. Pennington and J.~Portoles, \emph{{The Chiral Lagrangian parameters, l1,
  l2, are determined by the rho resonance}},
  \href{http://dx.doi.org/10.1016/0370-2693(94)01551-M}{\emph{Phys. Lett.}
  {\bfseries B344} (1995) 399--406}.

\bibitem{Vecchi:2007na}
L.~Vecchi, \emph{{Causal versus analytic constraints on anomalous quartic gauge
  couplings}},
  \href{http://dx.doi.org/10.1088/1126-6708/2007/11/054}{\emph{JHEP} {\bfseries
  11} (2007) 054}, [\href{https://arxiv.org/abs/0704.1900}{{\ttfamily
  0704.1900}}].

\bibitem{Manohar:2008tc}
A.~V. Manohar and V.~Mateu, \emph{{Dispersion Relation Bounds for pi pi
  Scattering}}, \href{http://dx.doi.org/10.1103/PhysRevD.77.094019}{\emph{Phys.
  Rev.} {\bfseries D77} (2008) 094019}.

\bibitem{Nicolis:2009qm}
A.~Nicolis, R.~Rattazzi and E.~Trincherini, \emph{{Energy's and amplitudes'
  positivity}}, \href{http://dx.doi.org/10.1007/JHEP05(2010)095,
  10.1007/JHEP11(2011)128}{\emph{JHEP} {\bfseries 05} (2010) 095},
  [\href{https://arxiv.org/abs/0912.4258}{{\ttfamily 0912.4258}}].

\bibitem{Bellazzini:2014waa}
B.~Bellazzini, L.~Martucci and R.~Torre, \emph{{Symmetries, Sum Rules and
  Constraints on Effective Field Theories}},
  \href{http://dx.doi.org/10.1007/JHEP09(2014)100}{\emph{JHEP} {\bfseries 09}
  (2014) 100}.

\bibitem{deRham:2017imi}
C.~de~Rham, S.~Melville, A.~J. Tolley and S.-Y. Zhou, \emph{{Massive Galileon
  Positivity Bounds}},
  \href{http://dx.doi.org/10.1007/JHEP09(2017)072}{\emph{JHEP} {\bfseries 09}
  (2017) 072}, [\href{https://arxiv.org/abs/1702.08577}{{\ttfamily
  1702.08577}}].

\bibitem{deRham:2017zjm}
C.~de~Rham, S.~Melville, A.~J. Tolley and S.-Y. Zhou, \emph{{UV complete me:
  Positivity Bounds for Particles with Spin}},
  \href{http://dx.doi.org/10.1007/JHEP03(2018)011}{\emph{JHEP} {\bfseries 03}
  (2018) 011}, [\href{https://arxiv.org/abs/1706.02712}{{\ttfamily
  1706.02712}}].

\bibitem{Mahoux:1969um}
G.~Mahoux and A.~Martin, \emph{{Extension of axiomatic analyticity properties
  for particles with spin, and proof of superconvergence relations}},
  \href{http://dx.doi.org/10.1103/PhysRev.174.2140}{\emph{Phys. Rev.}
  {\bfseries 174} (1968) 2140--2150}.

\bibitem{kotanski_transversity_1970}
A.~Kotański, \emph{Transversity amplitudes and their application to the study
  of particles with spin}, {\emph{Acta Phys Pol B1 45} (1970) }.

\bibitem{kotanski_kinematical_2016}
A.~Kotański, \emph{Kinematical singularities of the transversity amplitudes},
  \href{http://dx.doi.org/10.1007/BF02819831}{\emph{Il Nuovo Cimento A
  (1971-1996)} {\bfseries 56} (Jan., 2016) 737--754}.

\bibitem{deRham:2017xox}
C.~de~Rham, S.~Melville and A.~J. Tolley, \emph{{Improved Positivity Bounds and
  Massive Gravity}},
  \href{http://dx.doi.org/10.1007/JHEP04(2018)083}{\emph{JHEP} {\bfseries 04}
  (2018) 083}, [\href{https://arxiv.org/abs/1710.09611}{{\ttfamily
  1710.09611}}].

\bibitem{Giudice:2007fh}
G.~F. Giudice, C.~Grojean, A.~Pomarol and R.~Rattazzi, \emph{{The
  Strongly-Interacting Light Higgs}},
  \href{http://dx.doi.org/10.1088/1126-6708/2007/06/045}{\emph{JHEP} {\bfseries
  06} (2007) 045}, [\href{https://arxiv.org/abs/hep-ph/0703164}{{\ttfamily
  hep-ph/0703164}}].

\bibitem{Heisenberg:2014rta}
L.~Heisenberg, \emph{{Generalization of the Proca Action}},
  \href{http://dx.doi.org/10.1088/1475-7516/2014/05/015}{\emph{JCAP} {\bfseries
  1405} (2014) 015}, [\href{https://arxiv.org/abs/1402.7026}{{\ttfamily
  1402.7026}}].

\bibitem{deRham:2014wfa}
C.~de~Rham and R.~H. Ribeiro, \emph{{Riding on irrelevant operators}},
  \href{http://dx.doi.org/10.1088/1475-7516/2014/11/016}{\emph{JCAP} {\bfseries
  1411} (2014) 016}, [\href{https://arxiv.org/abs/1405.5213}{{\ttfamily
  1405.5213}}].

\bibitem{deRham:2016plk}
C.~de~Rham, A.~J. Tolley and S.-Y. Zhou, \emph{{The $\Lambda_{2}$ limit of
  massive gravity}},
  \href{http://dx.doi.org/10.1007/JHEP04(2016)188}{\emph{JHEP} {\bfseries 04}
  (2016) 188}, [\href{https://arxiv.org/abs/1602.03721}{{\ttfamily
  1602.03721}}].

\bibitem{Gabadadze:2017jom}
G.~Gabadadze, \emph{{Scale-up of $\Lambda_3$: Massive gravity with a higher
  strong interaction scale}},
  \href{http://dx.doi.org/10.1103/PhysRevD.96.084018}{\emph{Phys. Rev.}
  {\bfseries D96} (2017) 084018},
  [\href{https://arxiv.org/abs/1707.01739}{{\ttfamily 1707.01739}}].

\bibitem{deRham:2010ik}
C.~de~Rham and G.~Gabadadze, \emph{{Generalization of the Fierz-Pauli Action}},
  \href{http://dx.doi.org/10.1103/PhysRevD.82.044020}{\emph{Phys. Rev.}
  {\bfseries D82} (2010) 044020},
  [\href{https://arxiv.org/abs/1007.0443}{{\ttfamily 1007.0443}}].

\bibitem{deRham:2010kj}
C.~de~Rham, G.~Gabadadze and A.~J. Tolley, \emph{{Resummation of Massive
  Gravity}},
  \href{http://dx.doi.org/10.1103/PhysRevLett.106.231101}{\emph{Phys. Rev.
  Lett.} {\bfseries 106} (2011) 231101},
  [\href{https://arxiv.org/abs/1011.1232}{{\ttfamily 1011.1232}}].

\bibitem{Kotanski:1965zz}
A.~Kotanski, \emph{{Diagonalization Of Helicity Crossing Matrices}}, .

\bibitem{Bonifacio:2018vzv}
J.~Bonifacio and K.~Hinterbichler, \emph{{Bounds on Amplitudes in Effective
  Theories with Massive Spinning Particles}},
  \href{https://arxiv.org/abs/1804.08686}{{\ttfamily 1804.08686}}.

\bibitem{Burgess:1992gx}
C.~P. Burgess and D.~London, \emph{{Uses and abuses of effective Lagrangians}},
  \href{http://dx.doi.org/10.1103/PhysRevD.48.4337}{\emph{Phys. Rev.}
  {\bfseries D48} (1993) 4337--4351},
  [\href{https://arxiv.org/abs/hep-ph/9203216}{{\ttfamily hep-ph/9203216}}].

\bibitem{Kaplan:2005es}
D.~B. Kaplan, \emph{{Five lectures on effective field theory}},  2005.
\newblock \href{https://arxiv.org/abs/nucl-th/0510023}{{\ttfamily
  nucl-th/0510023}}.

\bibitem{Gripaios:2015qya}
B.~Gripaios, \emph{{Lectures on Effective Field Theory}},
  \href{https://arxiv.org/abs/1506.05039}{{\ttfamily 1506.05039}}.

\bibitem{Luty:2003vm}
M.~A. Luty, M.~Porrati and R.~Rattazzi, \emph{{Strong interactions and
  stability in the DGP model}},
  \href{http://dx.doi.org/10.1088/1126-6708/2003/09/029}{\emph{JHEP} {\bfseries
  09} (2003) 029}, [\href{https://arxiv.org/abs/hep-th/0303116}{{\ttfamily
  hep-th/0303116}}].

\bibitem{Nicolis:2004qq}
A.~Nicolis and R.~Rattazzi, \emph{{Classical and quantum consistency of the DGP
  model}}, \href{http://dx.doi.org/10.1088/1126-6708/2004/06/059}{\emph{JHEP}
  {\bfseries 06} (2004) 059},
  [\href{https://arxiv.org/abs/hep-th/0404159}{{\ttfamily hep-th/0404159}}].

\bibitem{deRham:2012ew}
C.~de~Rham, G.~Gabadadze, L.~Heisenberg and D.~Pirtskhalava,
  \emph{{Nonrenormalization and naturalness in a class of scalar-tensor
  theories}}, \href{http://dx.doi.org/10.1103/PhysRevD.87.085017}{\emph{Phys.
  Rev.} {\bfseries D87} (2013) 085017},
  [\href{https://arxiv.org/abs/1212.4128}{{\ttfamily 1212.4128}}].

\bibitem{deRham:2014zqa}
C.~de~Rham, \emph{{Massive Gravity}},
  \href{http://dx.doi.org/10.12942/lrr-2014-7}{\emph{Living Rev. Rel.}
  {\bfseries 17} (2014) 7}, [\href{https://arxiv.org/abs/1401.4173}{{\ttfamily
  1401.4173}}].

\bibitem{deRham:2016nuf}
C.~de~Rham, J.~T. Deskins, A.~J. Tolley and S.-Y. Zhou, \emph{{Graviton Mass
  Bounds}}, \href{http://dx.doi.org/10.1103/RevModPhys.89.025004}{\emph{Rev.
  Mod. Phys.} {\bfseries 89} (2017) 025004},
  [\href{https://arxiv.org/abs/1606.08462}{{\ttfamily 1606.08462}}].

\bibitem{Solomon:2017nlh}
A.~R. Solomon and M.~Trodden, \emph{{Higher-derivative operators and effective
  field theory for general scalar-tensor theories}},
  \href{http://dx.doi.org/10.1088/1475-7516/2018/02/031}{\emph{JCAP} {\bfseries
  1802} (2018) 031}, [\href{https://arxiv.org/abs/1709.09695}{{\ttfamily
  1709.09695}}].

\bibitem{Tasinato:2014eka}
G.~Tasinato, \emph{{Cosmic Acceleration from Abelian Symmetry Breaking}},
  \href{http://dx.doi.org/10.1007/JHEP04(2014)067}{\emph{JHEP} {\bfseries 04}
  (2014) 067}, [\href{https://arxiv.org/abs/1402.6450}{{\ttfamily 1402.6450}}].

\bibitem{Gripaios:2004ms}
B.~M. Gripaios, \emph{{Modified gravity via spontaneous symmetry breaking}},
  \href{http://dx.doi.org/10.1088/1126-6708/2004/10/069}{\emph{JHEP} {\bfseries
  10} (2004) 069}, [\href{https://arxiv.org/abs/hep-th/0408127}{{\ttfamily
  hep-th/0408127}}].

\bibitem{Heisenberg:2017hwb}
L.~Heisenberg, R.~Kase, M.~Minamitsuji and S.~Tsujikawa, \emph{{Black holes in
  vector-tensor theories}},
  \href{http://dx.doi.org/10.1088/1475-7516/2017/08/024}{\emph{JCAP} {\bfseries
  1708} (2017) 024}, [\href{https://arxiv.org/abs/1706.05115}{{\ttfamily
  1706.05115}}].

\bibitem{Heisenberg:2017xda}
L.~Heisenberg, R.~Kase, M.~Minamitsuji and S.~Tsujikawa, \emph{{Hairy
  black-hole solutions in generalized Proca theories}},
  \href{http://dx.doi.org/10.1103/PhysRevD.96.084049}{\emph{Phys. Rev.}
  {\bfseries D96} (2017) 084049},
  [\href{https://arxiv.org/abs/1705.09662}{{\ttfamily 1705.09662}}].

\bibitem{Chagoya:2016aar}
J.~Chagoya, G.~Niz and G.~Tasinato, \emph{{Black Holes and Abelian Symmetry
  Breaking}},
  \href{http://dx.doi.org/10.1088/0264-9381/33/17/175007}{\emph{Class. Quant.
  Grav.} {\bfseries 33} (2016) 175007},
  [\href{https://arxiv.org/abs/1602.08697}{{\ttfamily 1602.08697}}].

\bibitem{Kase:2017egk}
R.~Kase, M.~Minamitsuji and S.~Tsujikawa, \emph{{Relativistic stars in
  vector-tensor theories}},
  \href{http://dx.doi.org/10.1103/PhysRevD.97.084009}{\emph{Phys. Rev.}
  {\bfseries D97} (2018) 084009},
  [\href{https://arxiv.org/abs/1711.08713}{{\ttfamily 1711.08713}}].

\bibitem{ArkaniHamed:2002sp}
N.~Arkani-Hamed, H.~Georgi and M.~D. Schwartz, \emph{{Effective field theory
  for massive gravitons and gravity in theory space}},
  \href{http://dx.doi.org/10.1016/S0003-4916(03)00068-X}{\emph{Annals Phys.}
  {\bfseries 305} (2003) 96--118},
  [\href{https://arxiv.org/abs/hep-th/0210184}{{\ttfamily hep-th/0210184}}].

\bibitem{deRham:2010gu}
C.~de~Rham and G.~Gabadadze, \emph{{Selftuned Massive Spin-2}},
  \href{http://dx.doi.org/10.1016/j.physletb.2010.08.043}{\emph{Phys. Lett.}
  {\bfseries B693} (2010) 334--338},
  [\href{https://arxiv.org/abs/1006.4367}{{\ttfamily 1006.4367}}].

\bibitem{Bellazzini:2017fep}
B.~Bellazzini, F.~Riva, J.~Serra and F.~Sgarlata, \emph{{Beyond Positivity
  Bounds and the Fate of Massive Gravity}},
  \href{http://dx.doi.org/10.1103/PhysRevLett.120.161101}{\emph{Phys. Rev.
  Lett.} {\bfseries 120} (2018) 161101},
  [\href{https://arxiv.org/abs/1710.02539}{{\ttfamily 1710.02539}}].

\bibitem{deRham:2013qqa}
C.~de~Rham, L.~Heisenberg and R.~H. Ribeiro, \emph{{Quantum Corrections in
  Massive Gravity}},
  \href{http://dx.doi.org/10.1103/PhysRevD.88.084058}{\emph{Phys. Rev.}
  {\bfseries D88} (2013) 084058},
  [\href{https://arxiv.org/abs/1307.7169}{{\ttfamily 1307.7169}}].

\bibitem{Arkani-Hamed:2017jhn}
N.~Arkani-Hamed, T.-C. Huang and Y.-t. Huang, \emph{{Scattering Amplitudes For
  All Masses and Spins}},  \href{https://arxiv.org/abs/1709.04891}{{\ttfamily
  1709.04891}}.

\bibitem{Weinberg:1964ew}
S.~Weinberg, \emph{{Photons and Gravitons in s Matrix Theory: Derivation of
  Charge Conservation and Equality of Gravitational and Inertial Mass}},
  \href{http://dx.doi.org/10.1103/PhysRev.135.B1049}{\emph{Phys. Rev.}
  {\bfseries 135} (1964) B1049--B1056}.

\bibitem{vanDam:1970vg}
H.~van Dam and M.~J.~G. Veltman, \emph{{Massive and massless Yang-Mills and
  gravitational fields}},
  \href{http://dx.doi.org/10.1016/0550-3213(70)90416-5}{\emph{Nucl. Phys.}
  {\bfseries B22} (1970) 397--411}.

\bibitem{Zakharov:1970cc}
V.~I. Zakharov, \emph{{Linearized gravitation theory and the graviton mass}},
  {\emph{JETP Lett.} {\bfseries 12} (1970) 312}.

\end{thebibliography}\endgroup
%%%%%%%%%%%%%%%%%%%%%%%%%%%%

\end{document}